\documentclass[12pt,draftcls,onecolumn]{IEEEtran}
\usepackage{CJK}
\usepackage{mathptmx} 
\usepackage{latexsym,bm}
\usepackage{amsmath}
\usepackage{amsfonts}
 \usepackage{amssymb,amsmath}
 \usepackage{graphicx}
\usepackage{epstopdf}
\usepackage{epsfig}       
 \usepackage{amsfonts}
\usepackage{indentfirst}
\usepackage{graphicx}      
\usepackage{epsfig} 
\usepackage{epstopdf}
\usepackage{times}
\usepackage{bbm}
\usepackage[numbers,sort&compress]{natbib}
\usepackage{color}
\usepackage{diagbox}
\usepackage{makecell}
\allowdisplaybreaks
 \newtheorem{theorem}{Theorem}

\newtheorem{lemma}{Lemma}
\newtheorem{remark}{Remark}

\newtheorem{assumption}{Assumption}
\newtheorem{definition}{Definition}
\ifCLASSINFOpdf
\else
\fi
\begin{document}

\def\ba{\begin{array}}
\def\ea{\end{array}}
\def\ban{\begin{eqnarray*}}
\def\ean{\end{eqnarray*}}
\def\bd{\begin{description}}
\def\ed{\end{description}}
\def\be{\begin{equation}}
\def\ee{\end{equation}}
\def\bna{\begin{eqnarray}}
\def\ena{\end{eqnarray}}
%
\title{Cooperative Output Feedback Tracking Control of  Stochastic Linear Heterogeneous Multi-Agent Systems}
%
%
%

\author{Dianqiang~Li  and
        Tao~Li,~\IEEEmembership{Senior Member,~IEEE}
\thanks{*Corresponding author: Tao Li. This work was funded by the National Natural Science Foundation of
China under Grant No. 61977024.}
\thanks{D. Li is  with the School of Mathematical Sciences, East China Normal University, Shanghai
200241, China  (e-mail: ldq2015ahdx@126.com).}
\thanks{T. Li is with the Shanghai Key Laboratory of Pure Mathematics and Mathematical Practice, School of Mathematical Sciences, East China Normal University, Shanghai 200241, China (e-mail: tli@math.ecnu.edu.cn).}
}

%
%

\markboth{Journal of \LaTeX\ Class Files,~Vol.~14, No.~8, August~2015}%
{Shell \MakeLowercase{\textit{et al.}}: Bare Demo of IEEEtran.cls for IEEE Journals}
%



\maketitle

\begin{abstract}
We study cooperative output feedback tracking control of stochastic linear heterogeneous leader-following multi-agent systems. Each agent has a continuous-time linear  heterogeneous dynamics with  incompletely measurable state, and  there are additive and multiplicative noises along with  information exchange among agents. We propose a set of  admissible distributed  observation strategies for   estimating the leader's and the  followers' states, and  a set of admissible  cooperative output feedback control strategies based on the certainty equivalence principle.
By output regulation theory and stochastic analysis,  we show that for  observable  leader's dynamics and stabilizable and detectable followers' dynamics,
if the intensity coefficient of multiplicative noises multiplied by the sum of real parts of the leader' s unstable modes
is less than  $1/4$ of the minimum non-zero eigenvalue of graph Laplacian,
then   there exist  admissible distributed
observation and cooperative control strategies to ensure mean square bounded output tracking, provided the associated output regulation equations are solvable.
 Finally, the effectiveness of our control strategies is demonstrated by a numerical  simulation.
\end{abstract}


%
\IEEEpeerreviewmaketitle

\section{Introduction}
%
%
%
%
\IEEEPARstart{I}{n}
recent years,  distributed cooperative control  of multi-agent systems has attracted much attention by the system and control community and the research on the case with homogeneous dynamics has reached a reasonable degree of maturity
(\cite{Olfati-Saber,Ren213,Salehi,SuSZ2,ZhiyongYu}).

In practical applications, agents may have different dynamics.
For example, the differences in mass and orbits of satellites (\cite{Brouwer}),  velocities and mass of unmanned aerial vehicles (\cite{Murray2}), and generators and  loads of micro-grids (\cite{ Bevrani}) all lead to  dynamics of agents with different structures and parameters. Nowadays, many scholars have studied  distributed cooperative control problems of heterogeneous multi-agent systems (HMASs).
 As the dynamics of each agent is heterogeneous and even the  dimension of each agent's state is different,
  the cooperative output feedback control problem is more meaningful.
Wieland and Allg${\rm\ddot{o}}$wer (\cite{Wieland}) showed that
 the existence of a common internal model is   necessary  for output consensus under fixed topologies. By  the internal
model principle (\cite{Francis}),  Wieland and Allg${\rm\ddot{o}}$wer (\cite{Wieland1}) further studied output consensus  under time-varying topologies and  showed that the existence of a common internal model  is necessary and sufficient for  output  consensus if the  dynamics of each agent is  stabilizable and detectable.
 Assuming that only  output information can be transmitted among agents, Lunze (\cite{Lunze})
 proved that  the existence of a common internal model is   necessary  for output consensus.
By designing distributed observers and  decentralized laws, Grip \emph{et al}. (\cite{Grip}) investigated  output consensus  of HMASs with unmeasurable state.
By dividing the output regulator equations into  observable and unobservable parts,  Lewis \emph{et al}. (\cite{LewisF.L.66}) constructed a reduced-order synchronizer to achieve output consensus.
Inspired by classical output regulation theory (\cite{Francis1,Huang}),  Su and Huang (\cite{Su}) considered  cooperative output regulation of linear  HMASs by designing a distributed dynamic
feedback control law.
By exploiting properties of positive real transfer matrices, Alvergue \emph{et al}. (\cite{Alvergue}) proposed a output feedback control law to achieve output consensus.
Based on the solution of the output regulation equation, Yan \emph{et al}. (\cite{Yan43}) presented a distributed full information control law to achieve output consensus.
Huang \emph{et al}. (\cite{Huang74}) suggested an $H_{\infty}$ approach for cooperative output regulation of HMASs.
Yaghmaie \emph{et al}. (\cite{Yaghmaie74}) gave a linear matrix inequality condition for cooperative output regulation of HMASs.
By  a high-gain approach,
Meng \emph{et al}. (\cite{Meng74}) studied the output regulation of HMASs under deterministic switching topologies.
Based on the internal model principle and output regulation theory,  Kim \emph{et al}. \cite{Kim} and Su \emph{et al}. \cite{Su1} investigated the robust output regulation of linear  HMASs with parameter uncertainties.
Ding \emph{et al}. \cite{DingZT42} and Wang \emph{et al}. \cite{WangXH420} studied  the output regulation  of nonlinear HMASs.

Most of the above literature  assumed that each agent can get its neighbors'  information precisely. However, when each agent interacts with its neighbors through the communication network, communication processes are inevitably interfered by random noises due to uncertain communication environment.
 For example,  in  yaw  control of multiple unmanned aerial vehicles,  the yaw angles obtained from the  preceding vehicles through a communication network are usually interfered by random noises. For  multi-agent systems with additive noises,
sufficient conditions for mean square and almost sure consensus  were  given for discrete-time systems (\cite{Huang88,Aysal10,Kar0,Li11,Huang11,Huang55})
and continuous-time systems (\cite{Mac12,Hu59,ZhengYS23,LiW.Q.22,Liu22,Cheng22,LiuX.1,LiW.Q.,WuZ.H.,Cheng32}),  respectively.
Compared with additive noises, multiplicative noises may play a stabilizing  role in  almost sure stability  (\cite{Huang213}).
First-order continuous-time  multi-agent systems with multiplicative noises were studied in \cite{Ni22,Djaidja22,Li1,Zong2,Zong33}.
By the stochastic stability theorem and the generalized algebraic Riccati equation, Zong \emph{et al}. (\cite{Zong586}) studied  stochastic  consensus
of  continuous-time linear homogeneous  multi-agent systems.
Then the results were  generalized to the case with time delays in \cite{Zong5869}.

In this paper,
we investigate  cooperative output feedback tracking control of stochastic linear heterogeneous leader-following multi-agent systems.
 Each agent has a continuous-time linear  heterogeneous dynamics with  incompletely measurable state, and there are additive and multiplicative noises along with  information exchange among agents. We propose a set of  admissible distributed  observation strategies for   estimating each follower's own state and the leader's state, and  a set of admissible  cooperative output feedback control strategies based on the certainty equivalence principle.
 By output regulation theory and stochastic analysis,  we give sufficient conditions on the dynamics of agents, the network graph and the  noises for the existence of admissible distributed observation  and  cooperative control strategies to ensure mean square bounded output tracking.
The effectiveness of our control strategies is then demonstrated by a numerical simulation. The main contributions  are summarized as follows.

(i) Compared with the existing literature on HMASs, we assume that there are both additive and multiplicative noises along with information exchange among agents.  Multiplicative noises  make the estimate of the leader's state and the noises coupled together in a distributed
information structure. This leads to an additional diffusion term with coupled  estimates of the
leader's state and  network graphs in the  estimate error equation of the leader's state.
 To address this, firstly,  based on the duality principle  and  Lemma 3.1 in \cite{Zong586}, we give a sufficient condition for the existence of   positive define solution of the generalized Riccati equation related to the leader's dynamics. Then, we construct an appropriate stochastic Lyapunov function by  the inverse of this solution.
  Secondly,   by proving the negative definiteness of the quadratic form in the differential of this Lyapunov function multiplied by  an exponential function, we get the mean square upper bound of the estimate error for the leader's state.
Compared with  homogenous multi-agent systems with multiplicative measurement noises (\cite{Ni22,Djaidja22,Li1,Zong2,Zong33,Zong586}),  as  the dimensions of the followers' and  leader's state are different,
the method for analyzing
the tracking error equation for homogenous  systems is not applicable. To address this, we introduce an intermediate variable relying on the output regulation equation. By estimating the solution of the intermediate variable and  norm inequalities, we solve the mean square bounded
output tracking  of HMASs.

(ii) We show that for an observable  leader's dynamics and stabilizable and detectable followers' dynamics, if  the intensity coefficient of multiplicative noises multiplied by  the sum of real parts of unstable eigenvalues of the leader's dynamics
is less than  $1/4$ of the minimum non-zero eigenvalue of graph Laplacian,
then  there exist  admissible distributed observation and  cooperative control strategies to ensure mean square bounded output tracking, provided the associated output regulation equations are solvable.  Especially, if there are no additive measurement noises, then  there exist  admissible distributed observation and  cooperative control strategies to ensure mean square  output tracking.

(iii) For the case with one-dimensional   leader's  and  followers' dynamics,  we  give a necessary  and sufficient for the existence of admissible distributed observation  and cooperative control strategies without additive measurement noises to achieve mean square   output tracking under the star topology.

The rest of this paper is arranged as follows.
 Section II formulates the problem. Section III gives
the main results. Section IV gives a numerical
simulation to demonstrate the effectiveness of our control
laws. Section V concludes the paper.

\emph{Notation}: The symbol  $\mathbb{R}$ and  $\mathbb{R}_{+}$ denote real and nonnegative numbers, respectively;
$\mathbb{R}^{n}$ denotes the set of $n$-dimensional real column vectors;
$\mathbb{R}^{m\times n}$ denotes the set of
$m\times n$ dimensional real matrices; $\mathbf{0}_{N}$ represents the $N$-dimensional column vector with all zeros;
$\mathbf{1}_N$ denotes the $N$-dimensional
column vector with all ones; $I_{m}$ denotes the $m\times m$ dimensional identity matrix; $\mathrm{diag}\{A_1,\ldots ,A_N\}$ represents the block diagonal matrix  with entries being  $A_1,\ldots ,A_N$.
For a given vector or matrix $X$, $X^{\rm{T}}$ denotes its transpose, $\rm{Tr}(X)$ denotes its trace, and $\| X\|$ represents its  $2$-norm.
For a given real matrix $A\in\mathbb{R}^{n\times n}$, $\sigma(A)$ represents the spectrum of $A$, and
$\lambda_i(A)(i=1,\ldots,n)$ represents the $i$th eigenvalue of $A$ arranged in order of ascending real part. For a given complex number $Z$, $\mathbb{R}\mathbbm{e}(Z)$ represents its real part.
For a given real symmetric matrix $B\in\mathbb{R}^{n\times n}$, $\lambda_{\min}(B)$ is the minimum eigenvalue of $B$, and $\lambda_{\max}(B)$ is the maximum eigenvalue of $B$.
$A>0$ (or $A\geqslant0$) denotes  that $A$ is positive definite (or positive semi-definite) and $A<0$ (or $A\leqslant0$) denotes  that $A$ is negative definite (or negative semi-definite). For two real symmetric matrices $A$ and $B$, $A>B$ (or $A\geqslant B$) denotes that $A-B$ is  positive definite  (or $A-B$ is  positive semi-definite), and $A<B$(or $A\leqslant B$) denotes  that $A-B$ is  negative definite ( or $A-B$ is  negative semi-definite).
For two matrices $C$ and $D$, $C\otimes D$ denotes their Kronecker product.
Let
$(\Omega,\mathcal{F},\{\mathcal {F}_t\}_{t\geqslant t_0},\mathbb{P})$ a complete probability space with a filtration $\{\mathcal {F}_t\}_{t\geqslant t_0}$ satisfying the usual
conditions, namely, it is right continuous and increasing while $\mathcal {F}_0$ contains all $\mathbb{P}$-null sets;
$w(t)=\left(w_1(t),\ldots,w_m(t)\right)^T$ denotes a $m$-dimensional standard Brownian motion defined in $\left(\Omega,\mathcal {F},\left\{\mathcal{ F}_t\right\}_{t\geqslant t_0},\mathbb{P}\right)$. For a given random variable $X$, the mathematical expectation of $X$ is denoted by $\mathbb{E}[X]$.
The symbol $L^p\left([a,b];\mathbb{R}^{d}\right)$ denotes the family of $\mathbb{R}^{d}$-valued $\mathcal{F}_t$-adapted processes $\{f(t)\}_{a\leqslant t\leqslant b}$ such that $\int_a^b|f(t)|^p\textrm{d}t<\infty$ a.s.;
$\mathcal{L}^p(\mathbb{R}_{+};\mathbb{R}^{d})$ denotes the family of processes $\{f(t)\}_{t\geq0}$ such that for every $T>0$, $\{f(t)\}_{0\leqslant t\leqslant T}\in L^p\left([0,T];\mathbb{R}^{d}\right)$; $\mathcal{C}^{2,1}(\mathbb{R}^d\times \mathbb{R}_{+};\mathbb{R})$ denotes the family of all real valued functions $V(x,t)$ defined on $\mathbb{R}^d\times \mathbb{R}_{+}$, which are continuously twice differentiable in $x\in\mathbb{R}^d$ and once differentiable in $t\in\mathbb{R}_{+}$.

\section{Problem Formulations}
Consider a leader-following multi-agent system consisting of a leader and $N$ followers, where the leader is indexed by 0 and the $N$ followers are indexed by $1,\ldots,N$, respectively.
The dynamics of the leader  is given by
{\setlength\abovedisplayskip{6pt}
 \setlength\belowdisplayskip{6pt}
 \begin{equation}\label{b2}
\left\{ \begin{array}{l}
 {{\dot x}_0}(t) = {A_0}x_{0}(t),
 \\[0.5em]
 y_{0}(t) =C_0x_{0}(t), \\
\end{array} \right.
\end{equation}}%
where $x_0(t)\in \mathbb{R}^n$ is the state and $y_0(t)\in \mathbb{R}^p$ is the output of the leader, respectively; $A_0\in \mathbb{R}^{n\times n}$ and $C_0\in \mathbb{R}^{p\times n}$.

 The dynamics
of the $i$th follower is given by
{\setlength\abovedisplayskip{6pt}
 \setlength\belowdisplayskip{6pt}
 \begin{equation}\label{b1}
\left\{ \begin{array}{l}
 {{\dot x}_i} (t)= {A_i}x_{i}(t)+B_{i}u_{i}(t),
 \\[0.5em]
 y_{i}(t) =C_ix_{i}(t),\\
\end{array} \right.
\end{equation}}%
where $x_i(t)\in \mathbb{R}^{n_{i}}$ is the state, $u_i(t)\in \mathbb{R}^{m_i}$ is the input,
and $y_i(t)\in \mathbb{R}^p$ is the output of the $i$th follower, respectively; $A_i \in \mathbb{R}^{n_i\times n_i}$, $B_i\in \mathbb{R}^{n_i\times m_i}$, $C_i\in \mathbb{R}^{p\times n_i} $ and $p\leqslant m_i$.

We use $\mathcal{\overline{G}}=(\mathcal{\overline{V}},\mathcal{\overline{E}},\mathcal{\overline{A}})$ to represent a weighted graph formed by the leader and  $N$ followers,
and use  $\mathcal{G}=(\mathcal{V},\mathcal{E},\mathcal{A})$ to represent a subgraph formed by  $N$ followers, where the set of nodes $\mathcal{\overline{V}}=\left\{0,1,2\ldots,N\right\}$ and $\mathcal{V}=\mathcal{\overline{V}}\backslash\{0\}$, and the set of edges $\mathcal{\overline{E}}\subseteq \mathcal{\overline{V}}\times\mathcal{\overline{V}}$ and  $\mathcal{E}\subseteq \mathcal{V}\times\mathcal{V}$.  Denote the neighbors of the $i$th follower by $\mathcal{N}_i$. The adjacency matrix $\mathcal{A}=\left[a_{ij}\right]\in \mathbb{R}^{N\times N}$, $\mathcal{\overline{A}}=
\begin{bmatrix} 0& \mathbf{0}_{N}^{\rm T}\\a_0 & \mathcal{A}\end{bmatrix}\in \mathbb{R}^{(N+1)\times(N+1)}$, and  if $j\in \mathcal{N}_i$, then $a_{ij}=1$, otherwise $a_{ij}=0$;
$a=\left[a_{10},a_{20},\ldots,a_{N0}\right]^{\rm T}$ and if $0\in \mathcal{N}_i$, then $a_{i0}=1$, otherwise $a_{i0}=0$.
The Laplacian matrix  of  $\mathcal{\overline{G}}$ is given by $\mathcal{\overline{L}}=
\begin{bmatrix} 0& \mathbf{0}_{N}^{\rm T}\\-a &\mathcal{L}+F\end{bmatrix}\in \mathbb{R}^{(N+ 1)\times(N+1)}$, where $\mathcal{L}$ is the Laplacian matrix  of $\mathcal{G}$  and $F=diag\left(a_{10}, a_{20}, \ldots, a_{N0}\right)$.
\subsection{Admissible distributed observation  and cooperative control strategies}
 Since each agent has a  dynamics with  incompletely measurable state, we consider the following set of admissible  observation strategies
to estimate  agents' states. Denote
{\setlength\abovedisplayskip{6pt}
\setlength\belowdisplayskip{6pt}
\begin{eqnarray}
\mathcal{J}=\left\{J=\left\{(\Theta_i,\Xi_i),i=1,\ldots,N\right\}\right\},\nonumber
 \end{eqnarray}}%
where $\Theta_i$ represents an  observer of the $i$th follower to observe its own state, and $\Xi_i$ represents a distributed  observer of the $i$th follower to observe the leader's state. Here,
{\setlength\abovedisplayskip{1pt}
\setlength\belowdisplayskip{1pt}
\begin{eqnarray}\label{b3}
\Theta_i:
\dot {\hat{x}}_i(t) = {A_i}\hat{x}_{i}(t)+B_{i}u_{i}(t)+H_i\left(y_i(t)-C_i\hat{x}_i(t)\right),
 \end{eqnarray}}%
 where $x_i(t)$ is the state of the $i$th follower,  $\hat{x}_i(t)$ is the  estimate of $x_i(t)$,  and
 $H_i$ is the gain matrices to be designed.
{\setlength\abovedisplayskip{6pt}
\setlength\belowdisplayskip{6pt}
\begin{eqnarray}\label{b4}
\Xi_i:
\textrm{d}\hat{x}_{i0}(t)
=&&\hspace{-.6cm}A_0\hat{x}_{i0}(t)\textrm{d}t+G_{1i}\sum\limits_{j\in \mathcal{N}_i}a_{ij}\big[
C_0\left(\hat{x}_{j0}(t)
-\hat{x}_{i0}(t)\right)\textrm{d}t+\Upsilon_{ij}\textrm{d}w_{1ij}(t)\nonumber\\
&&\hspace{-.6cm}+\sigma_{ij}C_0\left(\hat{x}_{j0}(t)
-\hat{x}_{i0}(t)\right)\textrm{d}w_{2ij}(t)\big]
+G_{2i}a_{i0}\big[\left(y_{0}(t)-C_0\hat{x}_{i0}(t)\right)\textrm{d}t
\nonumber\\
&&\hspace{-.6cm}+\Upsilon_{i0}\textrm{d}w_{1i0}(t)+\sigma_{i0}\left( y_{0}(t)-C_0\hat{x}_{i0}(t)\right)\textrm{d}w_{2i0}(t)\big],
 \end{eqnarray}}%
where
$\hat{x}_{i0}(t)$ is the estimate of $x_0(t)$ by the $i$th follower; $\{w_{lij}(t), l=1,2,  i = 1, 2,\ldots ,N, j\in\mathcal{N}_i\}$  are   are  one dimensional standard Brownian motions,
 $\{\Upsilon_{ij}\in\mathbb{R}^{p},i=1,\ldots,N,  j\in\mathcal{N}_i\}$ and $\{\sigma_{ij}\in\mathbb{R}, i=1,\ldots,N, j\in\mathcal{N}_i\}$ represent   the intensity coefficient of additive  and  multiplicative measurement noises, respectively;
$G_{1i}$ and $G_{2i}$ are the gain matrices to be designed.

\vskip 0.2cm

\begin{remark}
For additive noises,  noise intensities  are independent of the system's state and for multiplicative noises,   noise intensities depend on the system's state.
Distributed consensus problems with additive and multiplicative
noises for continuous-time multi-agent systems have been studied in  \cite{Zong2,Wangt22,Wangt23}. Additive and multiplicative
noises co-exist in many  real systems.
      For example, the measurements by multiple sensors are often disturbed by both additive and multiplicative
noises in multi-sensor multi-rate systems (\cite{Fourati12}).
Here, the terms $\{\Upsilon_{ij}\textrm{d}w_{1ij}(t),i=1,\ldots,N,  j\in\mathcal{N}_i\}$ and $\{\sigma_{ij}C_0\left(\hat{x}_{j0}(t)
-\hat{x}_{i0}(t)\right)\textrm{d}w_{2ij}(t), i=1,\ldots,N, j\in\mathcal{N}_i\}$ in $(\ref{b4})$ represent  additive  and multiplicative measurement noises, respectively.
Compared with additive noises,
multiplicative noises  make the estimate of the leader's state and the noises coupled together in a distributed
information structure. This leads to an additional diffusion term with coupled  estimates of the
leader's state and  network graphs in the  estimate error equation of the leader's state.
 To address this, firstly,  based on the duality principle  and  Lemma 3.1 in \cite{Zong586}, we give a sufficient condition for the existence of   positive define solution $P$ of the generalized Riccati equation related to the leader's dynamics $(A_0,C_0)$ in Lemma 1. Then by $P^{-1}$, we construct an appropriate stochastic Lyapunov function.
 Secondly,   by proving that the quadratic form in the differential of the Lyapunov function multiplied by  an exponential function is
 negative definite, we get the mean square upper bound of the estimate error for the leader's state.
\end{remark}

\vskip 0.2cm
\begin{remark}
As a preliminary study, we assume that the measurement noises are Gaussian white noises.
In many  real systems,    noises can be considered as Gaussian white noises, and this assumption has been widely used in existing literature (\cite{Li11,Huang11,Hu59,Cheng22,LiW.Q.,Ni22}). For example, the measurement noises by multiple sensors are often modeled by Gaussian white noises in multi-sensor multi-rate systems (\cite{Fourati12}).
It would be interesting and challenging to investigate the case with non-Gaussian L\'{e}vy noises (\cite{KSchertzer,Bco2,KGaoi42}) in future.
\end{remark}
\vskip 0.2cm

\begin{remark}
We assume that the  leader's state matrix is known to all followers. This assumption has been widely used in the output regulation  of HMASs (\cite{Su,Alvergue,Kim,Karimi2}).
In fact, the assumption holds for many real systems, such as the position tracking of multiple wheeled mobile robots (\cite{Karimi2}).
\end{remark}

\vskip 0.2cm

For the output regulation problem of linear time-invariant systems, Huang \cite{Huang12} proposed a   state feedback control law
{\setlength\abovedisplayskip{1pt}
\setlength\belowdisplayskip{1pt}
\begin{eqnarray}\label{hjddc1}
u(t)=K_1x(t)+K_2\upsilon(t),
 \end{eqnarray}}%
where $x(t)$ is the state of the system, $\upsilon(t)$ is the state of the external system, and $K_1$ and $K_2$ are the gain matrices to be designed.

 We consider the following
set of admissible  distributed cooperative  control strategies based on the control law $(\ref{hjddc1})$ and the certainty equivalence principle
{\setlength\abovedisplayskip{1pt}
\setlength\belowdisplayskip{1pt}
\begin{eqnarray}
\mathcal{U}=\big\{U=\big\{u_i(t)=K_{1i}\hat{x}_i(t)+K_{2i}\hat{x}_{i0}(t), t\geqslant0,i=1,2,\ldots,N\big\}\big\}\nonumber
 \end{eqnarray}}%
and  the distributed control law of the $i$th follower is given by
{\setlength\abovedisplayskip{1pt}
\setlength\belowdisplayskip{1pt}
\begin{eqnarray}\label{1}
u_i(t)=K_{1i}\hat{x}_i(t)+K_{2i}\hat{x}_{i0}(t),
 \end{eqnarray}}%
where $\hat{x}_i(t)$ and $\hat{x}_{i0}(t)$ are given by $(\ref{b3})$ and $(\ref{b4})$, respectively, and $K_{1i}\in \mathbb{R}^{m_i\times n_i}$ and $K_{2i}\in \mathbb{R}^{m_i\times n}$ are the gain matrices to be designed.

\vskip0.2cm

\begin{remark}
Most of literature on  HMASs assumes that the state of each agent is known. However, in practical applications, due to cost constraints and other factors, the state of the system usually can't be obtained directly.
Compared with \cite{Huang12}, we use the state estimate $\hat{x}_i(t)$ and $\hat{x}_{i0}(t)$  instead of their true values in $(\ref{1})$  to design  the distributed control law.
For the  observers to estimate  the state of  leader, the follower who is not adjacent to the leader doesn't use  the leader's output $y_{0}(t)$,
but uses the relative estimate  of the leader's state
between the  follower and its neighbor $\hat{x}_{j0}(t)
-\hat{x}_{i0}(t)$, $j\in \mathcal{N}_i$. Therefore, the information structure of the observers are distributed.
\end{remark}

\vskip0.2cm
\subsection{Assumptions}
In this section, we formulate the assumptions on the agent's dynamics, the communication graph and the noises for the existence of admissible  distributed  observation and cooperative control strategies to achieve mean square  bounded output tracking.

For the dynamics of the leader and followers, we have the following assumptions.
\vskip0.2cm
\begin{assumption}\label{A1}
The pair $(A_i, B_i)$ is stabilizable, $i=1,2,\ldots,N$.
\end{assumption}
\vskip0.2cm
\begin{assumption}\label{A2}
The pair $(A_i, C_i)$ is detectable, $i=1,2,\ldots, N$.
\end{assumption}
\vskip0.2cm
\begin{assumption}\label{A3}
The pair $(A_0, C_0)$ is observable.
\end{assumption}
\vskip0.2cm
\begin{assumption}\label{A4}
The linear matrix equation
{\setlength\abovedisplayskip{1pt}
\setlength\belowdisplayskip{1pt}
\begin{eqnarray}\label{a1}
\begin{cases}
\Pi_iA_0=A_i\Pi_i+B_i\Gamma_i\\
C_i\Pi_i=C_0
\end{cases}
 \end{eqnarray}}%
has a solution $(\Pi_i, \Gamma_i)$ for each $i=1,2,\ldots, N$.
\end{assumption}

\vskip0.2cm
 \begin{remark} Note that there exists a solution $(\Pi_i, \Gamma_i)$ of
matrix equation $(\ref{a1})$ if and only if for all $\lambda\in\sigma(A_0)$,
{\setlength\abovedisplayskip{1pt}
\setlength\belowdisplayskip{1pt}
\begin{eqnarray}
\textrm{rank}\begin{bmatrix} \lambda-A_i& B_i\\C_i & 0\end{bmatrix}=n_i+p, \quad i=1,2,\ldots, N.\nonumber
 \end{eqnarray}}%
For more details, the readers may refer to  Theorem 1.9 in \cite{Huang12}.
\end{remark}
\vskip0.2cm

For  the noises and  the communication  graph, we have the following assumptions.
\begin{assumption}\label{A5}
The Brownian motions $\{w_{lij}(t),l=1,2,  i = 1, 2,\ldots ,N, j\in\mathcal{N}_i\}$  are  independent.
\end{assumption}

\vskip0.2cm

\begin{assumption}\label{A6}
The diagraph $\mathcal{\overline{G}}$ contains a spanning tree and the graph $\mathcal{G}$ is undirected.
\end{assumption}

\vspace{-1pt}
\section{Main results}
 Compared with \cite{Zong586},
we leave out $\mathbb{R}\mathbbm{e}(\lambda_n(A_0))\geqslant 0$ and develop the existence and uniqueness of positive define solution of the
generalized Riccati equation  related to the leader's dynamics in the following lemma.

\begin{lemma}\label{L1}
Suppose that  Assumption \ref{A3} holds. For any $\alpha\in\left[\lambda_0^{u}(A_0),\infty\right)$, where $\lambda_0^{u}(A_0)=\sum\limits_{i=1}^{n}\max\{\mathbb{R}\mathbbm{e}(\lambda_i(A_0)),0\}$,  the  generalized algebraic Riccati equation
{\setlength\abovedisplayskip{1pt}
\setlength\belowdisplayskip{1pt}
\begin{eqnarray}\label{a0}
A_0P+PA_0^{\rm T}-2\alpha PC_0^{\rm T}\left(I_p+C_0PC_0^{\rm T}\right)^{-1}C_0P+I_n=0
 \end{eqnarray}}%
has a unique positive solution $P$.
\end{lemma}
\vskip0.2cm
The proof is given in Appendix A.

\vskip0.2cm

%

\begin{definition}
 The leader-following HMASs $(\ref{b2})$$-$$(\ref{b1})$ under the distributed control law $(\ref{b3})$, $(\ref{b4})$ and $(\ref{1})$
  is said to achieve mean square bounded output tracking, if for any given initial values $x_0(0)$, $x_i(0)$,  $\hat{x}_i(0)$ and $\hat{x}_{i0}(0)$, $i=1,\ldots,N,$ there exists a constant $C>0$ such that
  {\setlength\abovedisplayskip{1pt}
\setlength\belowdisplayskip{1pt}
\begin{eqnarray}\label{chenZr24}
\limsup\limits_{t\rightarrow \infty}\mathbb{E}\left[\left\|y_i(t)-y_0(t)\right\|^2\right]\leqslant C,\quad i=1,2,\ldots,N.\nonumber
\end{eqnarray}}
  \end{definition}
  Especially, if $C=0$, then the leader-following HMASs $(\ref{b2})$$-$$(\ref{b1})$
   achieves mean square  output tracking.

\vskip0.2cm

Next, we will give  conditions  for the existence of admissible distributed observation  and cooperative control strategies to achieve mean square  bounded output tracking. Firstly, we define the mean square output tracking time for any given tracking precision. Denote  $t_{\varepsilon}=\inf\Big\{t: \sup_{s\geqslant t}\mathbb{E}\Big[\|y_i(s)-y_0(s)\|^2\Big]\leqslant\varepsilon, i=1,\ldots,N \Big\}$ for any given $\varepsilon>0$.  Denote $\sigma^2=\max\bigg\{\max\limits_{1\leqslant i,j\leqslant N} \sigma_{ij}^2$, $\max\limits_{1\leqslant i\leqslant N} \sigma_{i0}^2\bigg\}$.

\vskip0.2cm
\begin{theorem}\label{theorem1}
\textnormal{(I)}  Suppose that  Assumptions \ref{A1}$-$\ref{A6} hold and
$\sigma^2\lambda_0^{u}(A_0)<\frac{\lambda_1(\mathcal{L}+F)}{4}$. Then  there exists an admissible observation strategy $J\in\mathcal{J}$ and an admissible cooperative control strategy  $U\in\mathcal{U}$ such that   the HMASs $(\ref{b2})$$-$$(\ref{b1})$ achieve mean square bounded output tracking.

\textnormal{(II)}  Suppose that  Assumptions \ref{A1}$-$\ref{A6} hold and
$\sigma^2\lambda_0^{u}(A_0)<\frac{\lambda_1(\mathcal{L}+F)}{4}$.
 Choose $K_{1i}$ and $H_i, i=1,2,\ldots,N$ such that $A_i+B_iK_{1i}$ and $A_i-H_i C_i$ are Hurwitz, and choose $K_{2i}=\Gamma_i-K_{1i}\Pi_i$, $G_{11}=\ldots=G_{1N}=k_1PC_0^{\rm T}$ $\left(I_p+C_0PC_0^{\rm T}\right)^{-1}$, and $G_{21}=\ldots=G_{2N}=k_2PC_0^{\rm T}\left(I_p+C_0PC_0^{\rm T}\right)^{-1}$,
where $k_1$, $k_2\in(\underline{k},\overline{k})$,
$\underline{k}=\Big(\lambda_1(\mathcal{L}+F)-\sqrt{\lambda^2_1(\mathcal{L}+F)
-4\alpha\lambda_1(\mathcal{L}+F)\sigma^2}\Big)/2\lambda_1(\mathcal{L}+F)\sigma^2$,
$\overline{k}=\Big(\lambda_1(\mathcal{L}+F)+\sqrt{\lambda^2_1(\mathcal{L}+F)
-4\alpha\lambda_1(\mathcal{L}+F)\sigma^2}\Big)/2\lambda_1(\mathcal{L}+F)\sigma^2$,  $\alpha\in\big[\lambda_0^{u}(A_0),\lambda_0^{u}(A_0)+\varepsilon\big)$,
$\varepsilon\in\Big(0,$\\$\big(\lambda_1(\mathcal{L}+F)-4\sigma^2\lambda_0^{u}(A_0)\big)/4\sigma^2\Big)$,
then under the distributed control law $(\ref{b3})$, $(\ref{b4})$ and $(\ref{1})$,   the HMASs $(\ref{b2})$$-$$(\ref{b1})$ achieve mean square bounded output tracking and satisfy that
{\setlength\abovedisplayskip{1pt}
\setlength\belowdisplayskip{1pt}
\begin{eqnarray}\label{Z2l4}
&&\hspace{-.7cm}\limsup\limits_{t\rightarrow \infty}\mathbb{E}\left[\left\|y_i(t)-y_0(t)\right\|^2\right]
\leqslant\frac{6\rho_1^2\lambda^2_{\max}(P)\varpi_1^2\|P\|^2}{\rho_2^2}\left\| C_i\right\|^2\left\|B_iK_{2i}\right\|^2, i=1,\ldots,N,\nonumber
\end{eqnarray}}%
where $\varpi_1= \sum\limits_{j=1}^{N}
\Upsilon^{\rm T}_j\big[I_N\otimes G_1^{\rm T}P^{-1}G_1\big]\Upsilon_j+\Upsilon^{\rm T}_0\big[I_N\otimes G_2^{\rm T}P^{-1}G_2\big]\Upsilon_0$,  $\Upsilon_{j}= (\Upsilon^{\rm {T}}_{1j},\Upsilon^{\rm {T}}_{2j},\ldots,\Upsilon^{\rm {T}}_{Nj})^{\rm {T}}$,  $\Upsilon_{0}= (\Upsilon^{\rm {T}}_{10},\Upsilon^{\rm {T}}_{20},\ldots,\Upsilon^{\rm {T}}_{N0})^{\rm {T}}$,
 $(\Pi_i, \Gamma_i)$ is the solution  of
matrix equation $(\ref{a1})$,
 $P$ is the unique positive solution of  equation $(\ref{a0})$, $\rho_1$  and $\rho_2$ are positive constants satisfying $\left\|e^{(A_i+B_iK_{1i})t}\right\|\leqslant\rho_1 e^{-\rho_2 t}$.
\end{theorem}
\vskip0.2cm
The proof is given in Appendix B.

\vskip 0.2cm
We have the following theorem without additive measurement noises.

\begin{theorem}\label{theorem3} Suppose that  Assumptions \ref{A1}$-$\ref{A6}  hold and
$\sigma^2\lambda_0^{u}(A_0)<\frac{\lambda_1(\mathcal{L}+F)}{4}$. Then  there exists an admissible observation strategy $J\in\mathcal{J}$ and an admissible cooperative control strategy  $U\in\mathcal{U}$ such that  the HMASs $(\ref{b2})$$-$$(\ref{b1})$ achieve mean square output tracking. Especially,
choose the same $K_{1i}$, $H_i$, $K_{2i}$, $G_{1i}$ and $G_{2i}$, $i=1,\ldots,N$ as in Theorem 1-II,
then under the distributed control law $(\ref{b3})$, $(\ref{b4})$ and $(\ref{1})$, the leader-following HMASs $(\ref{b2})$$-$$(\ref{b1})$ achieve mean square output tracking.
The mean square output tracking time satisfies
{\setlength\abovedisplayskip{1pt}
  \setlength\belowdisplayskip{1pt}
   \begin{eqnarray}\label{lkvbjjuna1}
\left\{ \begin{array}{l}
t_{\varepsilon}\leqslant\max\left\{\frac{2}{\min\{\rho_2^2,\rho_4^2,\frac{1}{4\|P\|^2}\}},
\frac{\ln\left(\frac{\varpi_2}{\varepsilon}\right)}{2\min\left\{\rho_2,
\rho_4,\frac{1}{2\|P\|}\right\}}\right\},0<\varepsilon<\varpi_2,
 \\[0.5em]
t_{\varepsilon}=0, \quad \varepsilon\geqslant\varpi_2,
\end{array} \right.\nonumber
\end{eqnarray}}%
where
{\setlength\abovedisplayskip{1pt}
\setlength\belowdisplayskip{1pt}
\begin{eqnarray}
\varpi_2=&&\hspace{-.7cm}\Bigg\{2\rho_3^2\left\|C_i\right\|^2 \mathbb{E}\left[\left\|x_i(0)-\hat{x}_i(0)\right\|^2\right]+6\rho_1^2\left\|C_i\right\|^2\mathbb{E}\Big[\big\|\hat{x}_i(0)-\Pi_ix_0(0)\big\|^2\Big]\nonumber\\
&&\hspace{-.7cm}+\frac{6\rho_1^2\left\|C_i\right\|^2\left\|B_i K_{2i}\right\|^2\lambda_{\max}(P)}{\lambda_{\min}(P)}\mathbb{E}\left[\left
\|x_i(0)-\hat{x}_i(0)\right\|^2\right]\nonumber\\
&&\hspace{-.7cm}
+6\rho_1^2\rho_3^2 \left\|C_i\right\|^2\left\|H_i C_i\right\|^2
\mathbb{E}\left[\left\|x_i(0)-\hat{x}_i(0)\right\|^2\right]\Bigg\},\nonumber
\end{eqnarray}}%
 $(\Pi_i, \Gamma_i)$ is the solution  of
matrix equation $(\ref{a1})$,
 $P$ is the unique positive solution of  equation $(\ref{a0})$,  $\rho_1$, $\rho_2$, $\rho_3$ and $\rho_4$  are positive constants satisfying $\left\|e^{(A_i+B_iK_{1i})t}\right\|\leqslant\rho_1 e^{-\rho_2 t}$ and $\left\|e^{(A_i-H_i C_i)t}\right\|\leqslant\rho_3 e^{-\rho_4t}$, $i=1,\ldots N$.
\end{theorem}
\vskip0.2cm
The proof is given in Appendix B.
\vskip0.2cm
In the previous Theorem $\ref{theorem1}$$-$(II), we have given an upper bound of the mean square output tracking error.
Next, for  scalar systems, we  give a  lower bound of the mean square output tracking error under the star topology. We have the following assumptions.

\vskip 0.2cm
\begin{assumption}\label{Ak2}
The diagraph $\mathcal{\overline{G}}$ is a star topology, i.e. $a_{i0}=1$, $i = 1, 2,\ldots ,N$; $a_{ij}=0$, $i,j = 1, 2,\ldots ,N$.
\end{assumption}

\vskip0.2cm
\begin{assumption}\label{Ak1}
The Brownian motion $w_{2ij}(t), i = 1, 2,\ldots ,N, j\in\mathcal{N}_i$ and the  initial states of  the leader  and its observer $x_0(0)$, $\hat{x}_{i0}(0), i = 1, 2,\ldots,N,$ are  independent.
\end{assumption}

\vskip0.2cm

\begin{theorem}\label{theorem2} Consider the leader-following HMASs $(\ref{b2})$$-$$(\ref{b1})$, where $A_0=a_0\in\mathbb{R}$, $A_i=a_i\in\mathbb{R}$, $B_i=b_i\in\mathbb{R}$ and $C_i=c_i\in\mathbb{R}, i=1,\ldots,N$.
Suppose that $b_i\neq 0,$ $i=1,\ldots N$, $c_i\neq 0,$ $i=0,\ldots N$,  Assumption $\ref{Ak2}$ and Assumption $\ref{Ak1}$
  hold.
Choose $k$, $k_{1i}$ and $h_i, i=1,2,\ldots,N$ such that $a_0-kc_0<0$, $a_i+b_ik_{1i}<0$ and $a_i-h_i c_i<0$, and choose $k_{2i}=\gamma_i-k_{1i}\pi_i$, $G_{11}=\ldots=G_{1N}=0$, and $G_{21}=\ldots=G_{2N}=k$,
then under the distributed control law $(\ref{b3})$, $(\ref{b4})$ and $(\ref{1})$,  the closed-loop system satisfies that
{\setlength\abovedisplayskip{1pt}
 \setlength\belowdisplayskip{1pt}
 \begin{eqnarray}\label{vlpnl15}
\liminf\limits_{t\rightarrow \infty}\mathbb{E}\left[\left|y_i(t)-y_0(t)\right|^2\right]
\geqslant\frac{c_i^2 b_i^2 k_{2i}^2k^4\sigma_{i0}^2\Upsilon^2_{i0}c^2_0}{(a_0-kc_0)^2(a_i+b_ik_{1i})^2},\nonumber
 \end{eqnarray}}%
 where  $\pi_i=\frac{c_0}{c_i}$ and $\gamma_i=\frac{a_0c_0-a_ic_0}{b_ic_i}$.
\end{theorem}
\vskip0.2cm
The proof is given in Appendix B.
\vskip 0.2cm

In   Theorem  $\ref{theorem3}$, we have given the sufficient condition $\sigma^2\lambda_0^{u}(A_0)<\frac{\lambda_1(\mathcal{L}+F)}{4}$ for the existence of admissible distributed observation  and cooperative control strategies to achieve mean square  output tracking.
Next, for  scalar  systems, we give the necessary and sufficient condition under the star topology.
\vskip 0.2cm
\begin{theorem}\label{theorem4} Consider the same scalar  systems as in Theorem $\ref{theorem2}$.
Suppose that $b_i\neq 0,$ $i=1,\ldots N$, $c_i\neq 0,$ $i=0,\ldots N$,  and Assumptions $\ref{Ak2}$$-$$\ref{Ak1}$
  hold.
Then  there exists an admissible observation strategy $J\in\mathcal{J}$ and an admissible cooperative control strategy  $U\in\mathcal{U}$ such that  the HMASs $(\ref{b2})$$-$$(\ref{b1})$ achieve mean square  output tracking if and only if $\sigma^2a_0<\frac{1}{2}$, where $\sigma^2= \max\limits_{1\leqslant i\leqslant N} \sigma_{i0}^2$.
\end{theorem}

The proof is given in Appendix B.
\vskip 0.2cm

  \begin{remark} Theorem $\ref{theorem4}$   shows that if the multiplicative noises are sufficiently strong or the dynamics of leader is sufficiently unstable, i.e.  $\sigma^2a_0\geqslant\frac{1}{2}$,
then the followers  can't achieve mean square output tracking for  any  distributed observation and cooperative control strategies. In fact, $\lambda_0^{u}(A_0)=a_0$ and $\lambda_1(\mathcal{L}+F)=a_{i0}=1$ under  Assumption 7 in Theorem $\ref{theorem4}$.
The condition $\sigma^2\lambda_0^{u}(A_0)<\frac{\lambda_1(\mathcal{L}+F)}{4}$ in Theorem  $\ref{theorem3}$ degenerates to $\sigma^2a_0<\frac{1}{4}$ for the case in Theorem $\ref{theorem4}$.
Here, the gap between $\sigma^2a_0<\frac{1}{2}$ and  $\sigma^2a_0<\frac{1}{4}$ shows the conservativeness of the condition
  $\sigma^2\lambda_0^{u}(A_0)<\frac{\lambda_1(\mathcal{L}+F)}{4}$.
\end{remark}

\vskip 0.2cm
\begin{remark}
The condition $\sigma^2\lambda_0^{u}(A_0)<\frac{\lambda_1(\mathcal{L}+F)}{4}$ in  Theorem  $\ref{theorem3}$ together with the condition $\sigma^2a_0<\frac{1}{2}$ in Theorem  $\ref{theorem4}$ shows   the influence of multiplicative noises, the leader's dynamics and the communication graph on the existence of admissible distributed observation  and cooperative control strategies to achieve mean square  output tracking.
It is shown that  smaller
  multiplicative  noises, more stable leader's dynamics and more connected communication graphs are all more helpful for
 the cooperatability of the system. This is consistent with intuition. Theorem  $\ref{theorem4}$ shows that if the noise intensity coefficient $\sigma^2$ is sufficiently large, then the mean square  output tracking can't be achieved,   even if the communication graph has a spanning tree, i.e. $\lambda_1(\mathcal{L}+F)>0$. This is totally different from the noise-free case, which implies that  multiplicative noises indeed have an essential impact on the cooperatability of stochastic multi-agent systems.
\end{remark}

\section{Numerical  simulation}
In this section, we will use a numerical  example to demonstrate
the effectiveness of our control laws.

\emph{Example} 4.1.
We  consider  a heterogeneous fleet
consisting of a leader aircraft and three follower aircrafts, and demonstrate that the sideslip angle  of  followers can track  that of the leader under the distributed control law $(\ref{b3})$, $(\ref{b4})$ and $(\ref{1})$.

The leader is a Lockheed L-1011, the first follower  is a Boeing-767, the second follower  and the
 third follower are   McDonnell Douglas F/A-18/HARV fighters.
 Referring to \cite{Tomashevich},
 the dynamics of the leader is given by $(\ref{b2})$,
where $x_0(t)\in \mathbb{R}^3$ and $y_0(t)\in \mathbb{R}$;
the  components of  $x_0(t)$ are  sideslip angle, roll angle,
roll rate, respectively; $A_0=\begin{bmatrix}-0.1170&0.0386&-0.0003\\0&0&1\\-5.200&0&-1\end{bmatrix}$ and $C_0=\begin{bmatrix}1&0&0\end{bmatrix}$.
 The dynamics
of the $i$th follower is given by $(\ref{b1})$,
where $x_i(t)\in \mathbb{R}^4$ and $y_i(t)\in \mathbb{R}$, $i=1,2,3$;
the  components of  $x_i(t)$ are  sideslip angle, roll angle,
roll rate and yaw rate, respectively;
$A_1=\begin{bmatrix}-0.1245&0.0414&0.0350&-0.9962\\0&0&1&0.0357\\-15.2138&0.0032&-2.0587&0.6458\\1.6447&-0.0022&-0.0447&-0.1416\end{bmatrix}$,

  $A_2=A_3=\begin{bmatrix}-0.1703&0.0440&0.0490&0.9980\\0&0&1&0.0491\\-15.5763&0&-2.3142&0.5305\\3.0081&0&-0.0160&-0.1287\end{bmatrix}$,
$B_1=\begin{bmatrix}-0.0049&0.0237\\0&0\\-4.0379&0.9613\\-0.0568&-1.2168\end{bmatrix}$,
$B_2=B_3=\begin{bmatrix}-0.0069&-0.0153&0.0380\\0&0&0\\23.3987&21.4133&3.2993\\-0.1644&0.3313&-1.9836\end{bmatrix}$,
 $C_1=C_2=C_3=\begin{bmatrix}1&0&0&0\end{bmatrix}$.

The communication topology $\mathcal{\overline{G}}=(\mathcal{\overline{V}},\mathcal{\overline{E}},\mathcal{\overline{A}})$ is shown in \text{ Fig. 1}, where $\mathcal{\overline{A}}=
\begin{bmatrix} 0&0& 0&0\\1 & 0&0&0\\1&0&0&1\\0&0&1&0\end{bmatrix}$.
By \text{ Fig. 1}, we get $\lambda_2(\mathcal{\overline{L}})$$=\lambda_1(\mathcal{L}+F)=\frac{3-\sqrt{5}}{2}$.
The additive   measurement noises in $(\ref{b4})$ are given by
  $0.9\textrm{d}w_{1ij}(t),i=1,2,3,  j\in\mathcal{N}_i$.
 The  multiplicative measurement noises in $(\ref{b4})$ are given by
  $1.2C_0\left(\hat{x}_{j0}(t)-\hat{x}_{i0}(t)\right)\textrm{d}w_{2ij}(t), i=1,2,3, j\in\mathcal{N}_i$.
The initial states of agents are given by $x_0(0)=\begin{bmatrix}0.2&0.1&0.2\end{bmatrix}^{\rm T}$, $x_1(0)=\begin{bmatrix}-0.5&0.1&0.2&0.1\end{bmatrix}^{\rm T}$,  $\hat{x}_1(0)=\big[0.1\quad 0.3\quad 0.1\quad 0.2\big]^{\rm T}$, $\hat{x}_{10}(0)=\begin{bmatrix}-0.5&0.1&-0.1\end{bmatrix}^{\rm T}$, $x_2(0)=\begin{bmatrix} -0.1&0.1&0.2&0.1\end{bmatrix}^{\rm T}$
$\hat{x}_2(0)=  \begin{bmatrix}-0.2&0.2&0.1&0.3\end{bmatrix}^{\rm T}$, $\hat{x}_{20}(0)=\begin{bmatrix}-0.2&0.1&0.2\end{bmatrix}^{\rm T}$, $x_3(0)=\begin{bmatrix}0.4&-0.2&0.1&0.3\end{bmatrix}^{\rm T}$,  $\hat{x}_3(0)=\begin{bmatrix}0.1&0.2&0.1&0.1\end{bmatrix}^{\rm T}$ and $\hat{x}_{30}(0)=\begin{bmatrix}0.3&0.2&0.2\end{bmatrix}^{\rm T}$.
\begin{figure}
\centering
\includegraphics[height=2in, width=3in]{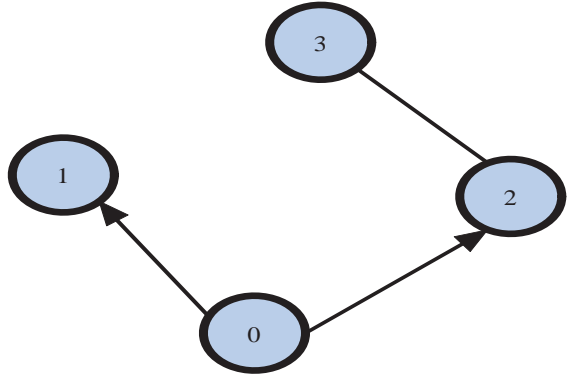}
\caption{{The communication topology graph.}}
\end{figure}

\begin{figure}
\begin{minipage}[t]{0.5\linewidth}
\centering
\includegraphics[height=2in, width=3in]{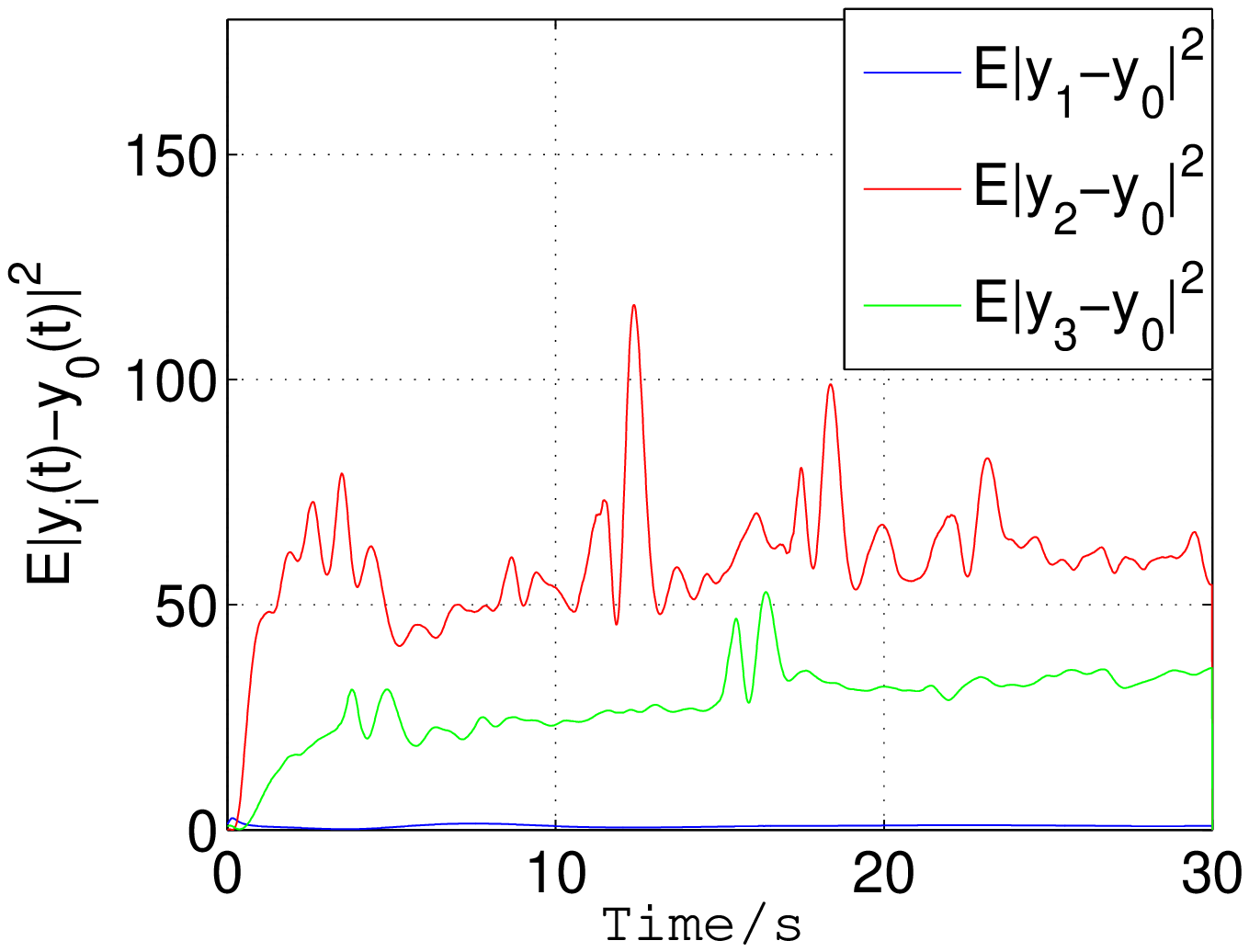}
\caption{Mean square output tracking errors.}
\end{minipage}%
\begin{minipage}[t]{0.5\linewidth}
\centering
\includegraphics[height=2in, width=3in]{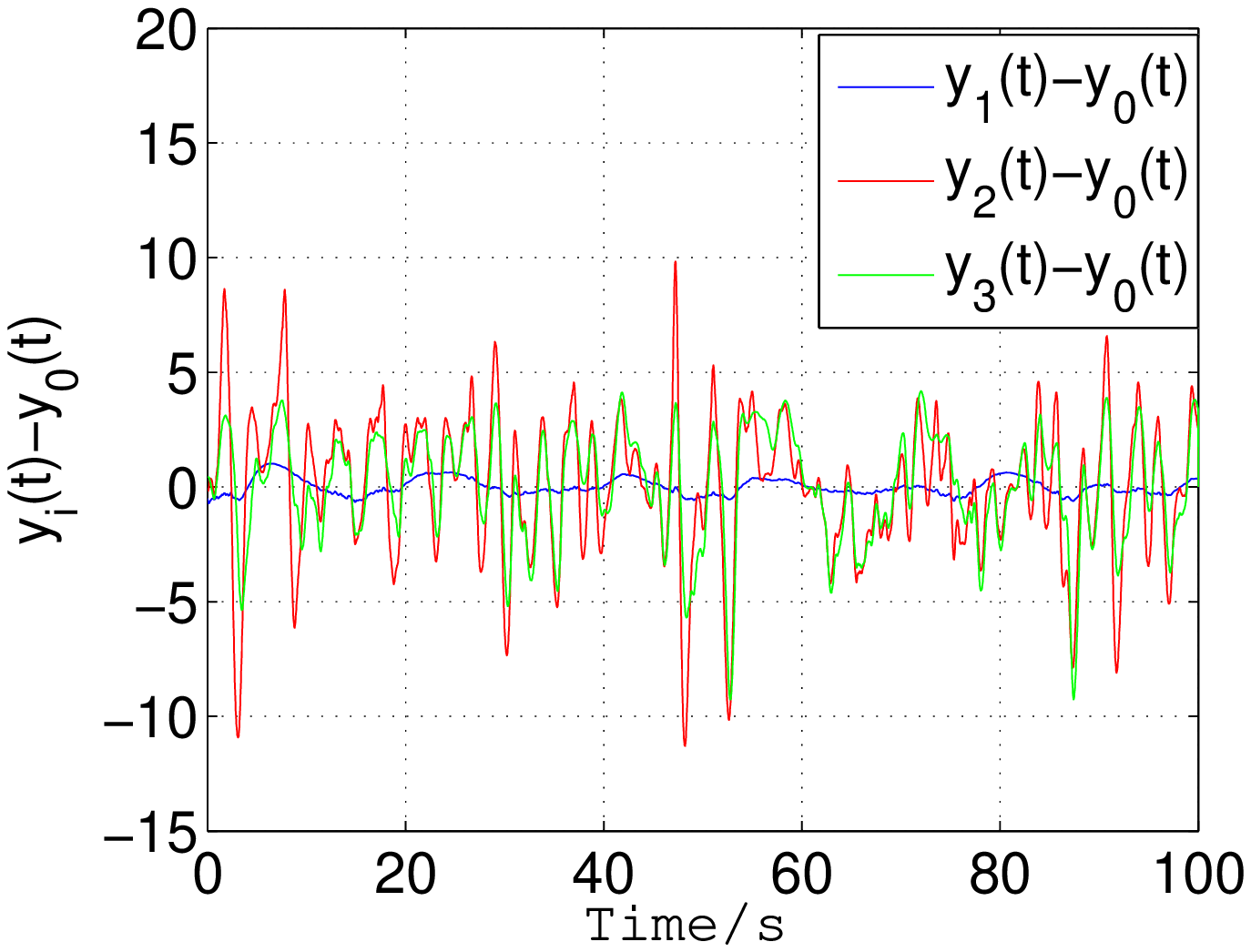}
\caption{ Sample paths of output tracking errors.}
\end{minipage}%
\end{figure}

\begin{figure}
\begin{minipage}[t]{0.5\linewidth}
\centering
\includegraphics[height=2in, width=3in]{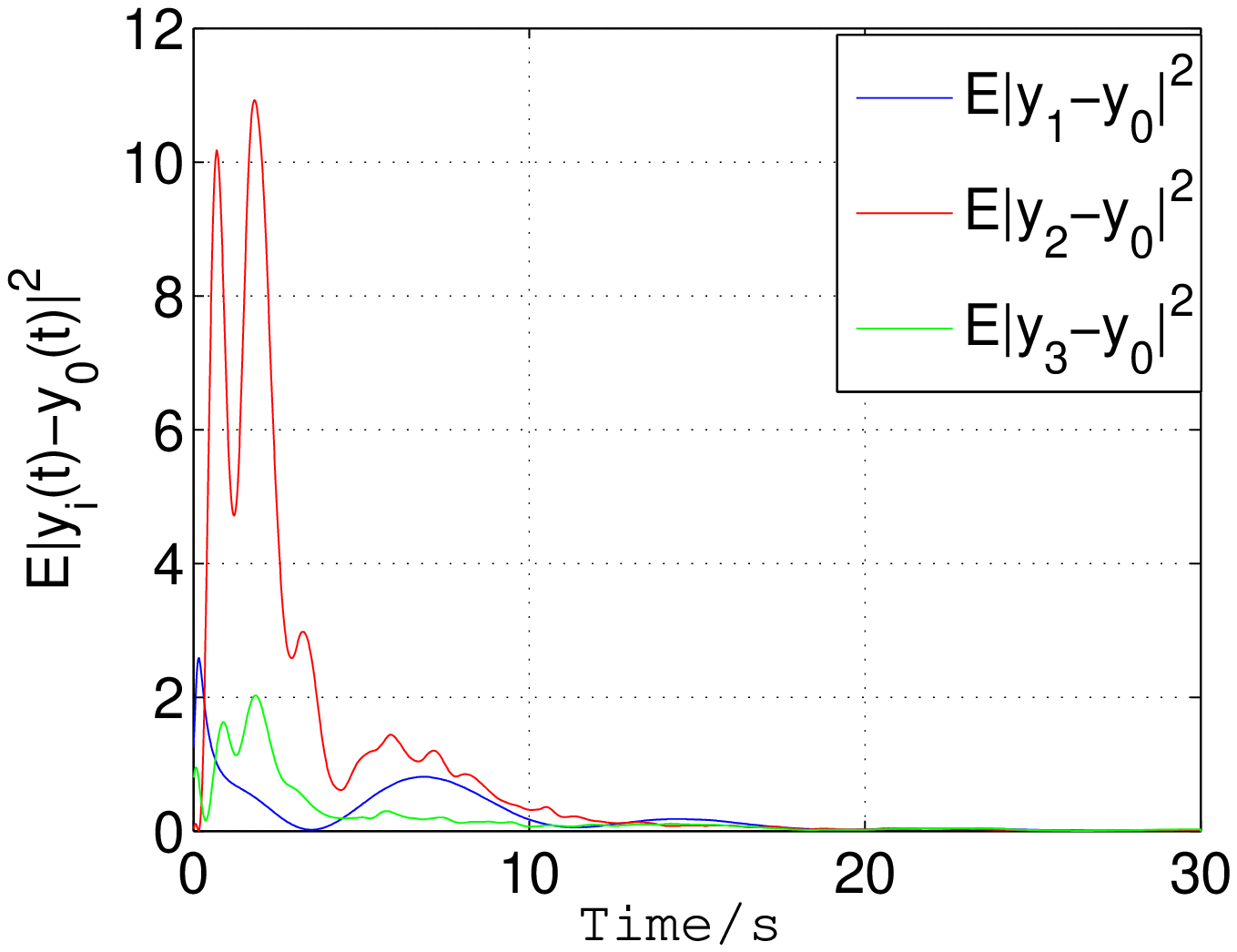}
\caption{Mean square output   tracking errors without\protect\\ additive  noises.}
\end{minipage}%
\begin{minipage}[t]{0.5\linewidth}
\centering
\includegraphics[height=2in, width=3in]{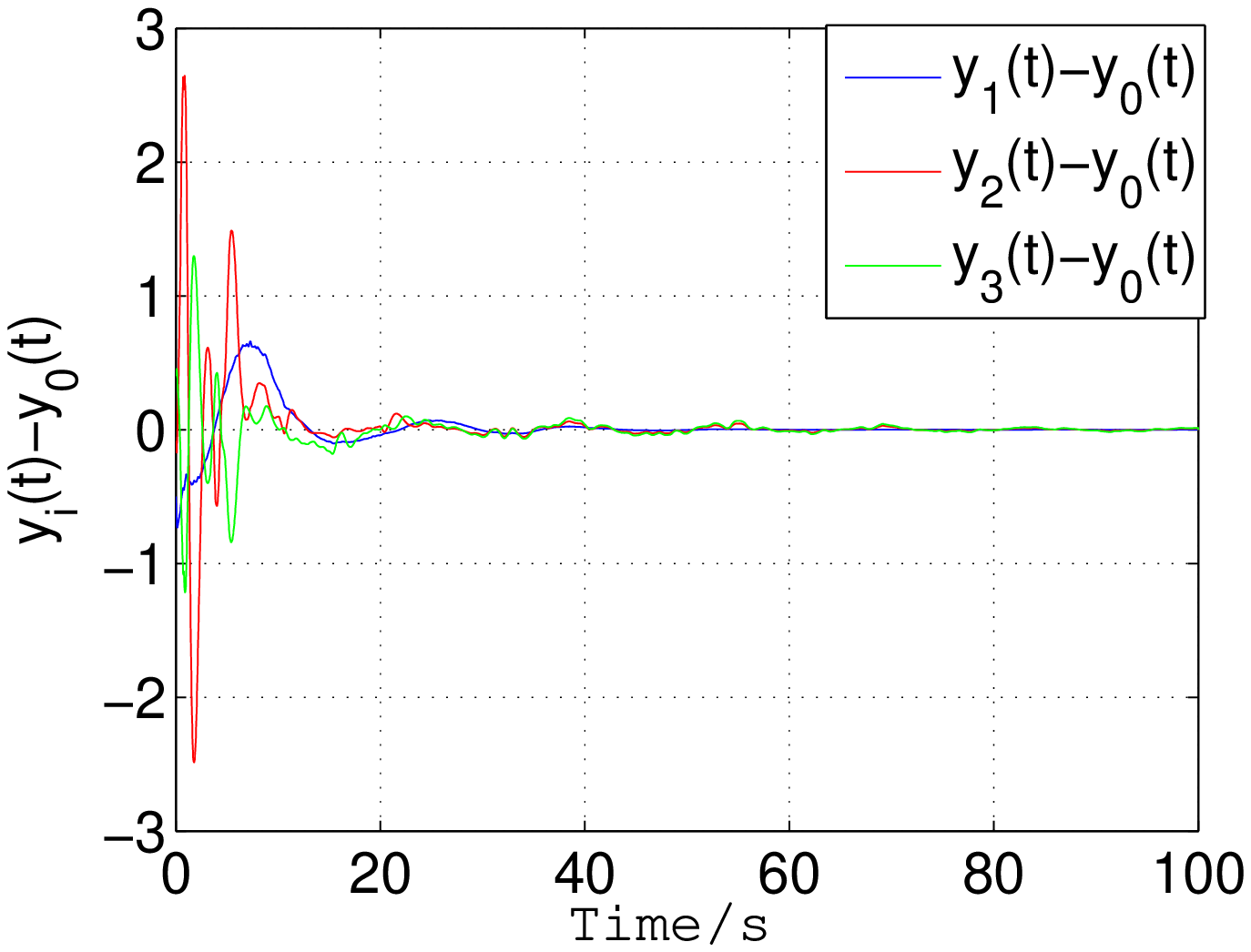}
\caption{Sample paths of output tracking errors without  additive  noises.}
\end{minipage}%
\end{figure}

 It can be verified that the pair $(A_i,B_i)$ is controllable for $i=1,2,3$, and the pair $(C_i,A_i)$ is observable for $i=0,1,2,3$.

Choose
$K_{11}=\begin{bmatrix}-10&8 &6 &12\\-10& -3& -6 &-3\end{bmatrix}$,
$K_{12}=K_{13}=\begin{bmatrix}-12& -10& -2& -3\\1& -2& -1& -1\\5& -1& 0& 1\end{bmatrix}$,
$H_1=\begin{bmatrix}-2\\-1\\2\\1\end{bmatrix}$,
$H_2=H_3=\begin{bmatrix}-5\\-3\\-4\\-3\end{bmatrix}$
such that $A_i+B_iK_{1i}$ and $A_i+H_{i}C_i$ are Hurwitz for $i=1,2,3$.
By $(\ref{a1})$, we have

$\Pi_1=\begin{bmatrix}1&0&0\\2.8072& -1.4746& -2.2583\\11.3945& 0.1118& 0.7853\\0.5721& -0.0952& -0.0668\end{bmatrix}$,
$\Pi_2=\Pi_3=\begin{bmatrix}1& 0& 0\\36.6879& 14.2062&  10.8173\\-60.6086& 1.4485& 3.4095\\1.3468& -0.6584& -0.6428\end{bmatrix}$,
$\Gamma_1=\begin{bmatrix}-7.9039& -0.1823& 0.1666\\1& 0& 0\end{bmatrix}$,
$\Gamma_2=\Gamma_3=\begin{bmatrix}-6.0891& 0.0582& 0.2688\\1& 0& 0\\1& 0& 0\end{bmatrix}$.
Since $\lambda_0^{u}(A_0)=$\\$\sum\limits_{i=1}^{3}\max\{\mathbb{R}\mathbbm{e}(\lambda_i(A_0)),$ $0\}=0.0486$, we  select $\alpha=0.65$, $k_1=0.38$ and $k_2=0.3$.

The mean square errors  of sideslip angle and  sample paths of sideslip angle errors   under the distributed control law $(\ref{b3})$, $(\ref{b4})$ and $(\ref{1})$ are shown in \text{ Fig. 2} and \text{ Fig. 3}.
If there are no additive measurement noises, i.e. $\Upsilon_{01}=\Upsilon_{02}=\Upsilon_{23}=\Upsilon_{32}=0$, then mean square errors  of sideslip angle and  sample paths of sideslip angle errors
are shown in \text{ Fig. 4} and \text{ Fig. 5}.  

\text{ Fig. 2} together with  \text{Fig. 4} shows that additive measurement noises lead to a non-zero  mean square
tracking error while multiplicative measurement noises have no impact on the mean square
tracking error.
\text{ Fig. 5} shows that multiplicative measurement noises make the sample paths of sideslip angle errors fluctuate greatly at the beginning and then vanish.
\text{ Fig. 3} shows that additive measurement noises make the sample paths of sideslip angle errors fluctuate during the entire process.

 \section{Conclusion}
  In this paper, we have studied  cooperative output feedback tracking control of stochastic linear heterogeneous leader-following multi-agent systems.  By output regulation theory and stochastic analysis,  we have shown that for  observable  leader's dynamics and stabilizable and detectable followers' dynamics, if (i) the associated output regulation equations are solvable,
  (ii)    the intensity coefficient of multiplicative noises multiplied by  the sum of real parts of unstable eigenvalues of the leader's dynamics
is less than  $1/4$ of the minimum non-zero eigenvalue of graph Laplacian,
then   there exist  admissible distributed
observation and cooperative control strategies based on the certainty equivalent principle to ensure mean square bounded output tracking.
Especially, if there are no additive measurement noises, then   there exist  admissible distributed
observation and cooperative control strategies  to achieve mean square output tracking.
 There are still many other interesting topics to be studied in  future.  Efforts can be made to investigate the consensus problem of HMASs under Markovian switching topologies and HMASs with delays.

\appendices
\section{Definitions and Lemmas}
\begin{definition} \textnormal{(\cite{Mao33}, It\^{o}'s formula)}\label{L3}  Let $x(t)$ be a $d$-dimensional It\^{o} process on $t\geq 0$ with the stochastic differential
   \begin{eqnarray}
dx(t)= f(t)dt+g(t)\textrm{d}w(t),\nonumber
   \end{eqnarray}
where $f\in \mathcal{L}^1(\mathbb{R}_{+};\mathbb{R}^d)$ and $g\in \mathcal{L}^2(\mathbb{R}_{+};\mathbb{R}^{d\times m})$. Let  $V \in C^{2,1}(\mathbb{R}^d\times \mathbb{R}_{+};\mathbb{R})$. Then
$V(x(t),t)$ is also an It\^{o} process with the stochastic differential given by
 \begin{eqnarray}
\textrm{d}V(x(t),t)
= &&\hspace{-.7cm}\left[V_t(x(t),t)+V_x(x(t),t)f(t)+\frac{1}{2}Tr(g^{\rm T}(t)V_{xx}(x(t),t)g(t))\right]\textrm{d}t+V_x(x(t),t)g(t)\textrm{d}w(t) \nonumber.
   \end{eqnarray}
\end{definition}

\vspace{0.2cm}

\emph{Proof of Lemma \ref{L1}}:
By Assumption \ref{A3}, we know that $\left(A_0^{\rm T},C_0^{\rm T}\right)$ is controllable. If $\mathbb{R}\mathbbm{e}(\lambda_n(A_0))\geqslant0$ holds, we choose $\alpha>\lambda_0^{u}(A_0)$.  By Lemma 3.1 in \cite{Zong586},  we know that $A^{\rm T}P+PA-2\alpha PB\left(I_p+B^{\rm T}PB\right)^{-1}$\\$B^{\rm T}P+I_n=0$ has a unique positive solution $P$, where $A=A_0^{\rm T}$ and $B=C_0^{\rm T}$. If $\mathbb{R}\mathbbm{e}(\lambda_n(A_0))<0$ holds, we choose $\alpha=\lambda_0^{u}(A_0)=0$. Therefore, we know that the equation  $(\ref{a0})$ is a Lyapunov equation and it has a unique positive solution $P$.
$ \hfill\square$
\vspace{0.2cm}
\begin{lemma} \textnormal{(\cite{LiW.Q.})}\label{L2}
If $A$ is Hurwtiz, then the solution of the system
\begin{eqnarray}
\textrm{d}x(t)=Ax(t)\textrm{d}t+B\textrm{d}w(t)\nonumber
 \end{eqnarray}%
 satisfies
    \begin{eqnarray}
\lim_{t\rightarrow \infty}\mathbb{E}\left[\left\| x(t)\right\|^2\right]={\rm{Tr}}\left\{\int_{0}^{\infty}e^{A s}BB^Te^{A^Ts}\textrm{d}s\right\}.\nonumber
 \end{eqnarray}%
\end{lemma}

\vspace{0.2cm}
 \begin{lemma}\label{L66}
Denote the matrix $S_{2ij}=[s_{kl}]_{N\times N}$  satisfying
\begin{eqnarray}
\left\{ \begin{array}{l}
s_{kl}=-a_{ij},\quad  k=i, l=i,
 \\[0.5em]
s_{kl}=a_{ij}, \quad k=i,  l=j,
 \\[0.5em]
 s_{kl}=0, \quad  otherwise,
\end{array} \right.\nonumber
\end{eqnarray}%
 then
$\sum\limits_{i,j=1}^{N}\left({S_{2ij}^{\rm T}}S_{2ij}\right)=2\mathcal{L}$, where $\mathcal{L}$ is the Laplacian matrix  of $\mathcal{G}$, $\mathcal{A}$ is the adjacency matrix of $\mathcal{G}$
 and $a_{ij}$ is the element of $\mathcal{A}$.
\end{lemma}

\emph{Proof :}
 By the definition of $S_{2ij}$, $1\leqslant i\leqslant j\leqslant N$, we have
 {\setlength\abovedisplayskip{2pt}
\setlength\belowdisplayskip{2pt}
  \begin{eqnarray}
  S_{2ij}=
\bordermatrix{%
&1  &\quad &i&\quad&j&\quad\cr
1 &0&\cdots&0&\cdots&0&\cdots\cr
\quad&\vdots&\cdots&\vdots&\cdots&\vdots&\cdots\cr
i &0&\cdots&-a_{ij}&\cdots&a_{ij}&\cdots\cr
&\vdots&\cdots&\vdots&\cdots&\vdots&\cdots\cr
j&0&\cdots&0&\cdots&0&\cdots\cr
&\vdots&\cdots&\vdots&\cdots&\vdots&\cdots\cr
}_{N\times N},\nonumber
\end{eqnarray}}%
where the omitted elements are zero.

By the above equation, we obtain
{\setlength\abovedisplayskip{2pt}
\setlength\belowdisplayskip{2pt}
  \begin{eqnarray}\label{kqq2}S_{2ij}^{\rm T}S_{2ij}=\bordermatrix{%
&1  &\quad &i&\quad&j&\quad\cr
1 &0&\cdots&0&\cdots&0&\cdots\cr
\quad&\vdots&\cdots&\vdots&\cdots&\vdots&\cdots\cr
i &0&\cdots&a_{ij}^2&\cdots&-a_{ij}^2&\cdots\cr
&\vdots&\cdots&\vdots&\cdots&\vdots&\cdots\cr
j&0&\cdots&-a_{ij}^2&\cdots&a_{ij}^2&\cdots\cr
&\vdots&\cdots&\vdots&\cdots&\vdots&\cdots\cr
}_{N\times N}.\nonumber
\end{eqnarray}}%
By $a_{ij}=0$ or $a_{ij}=1$, we have $a_{ij}^2=a_{ij}$,
which together with the above equation gives
{\setlength\abovedisplayskip{2pt}
\setlength\belowdisplayskip{2pt}
  \begin{eqnarray}S_{2ij}^{\rm T}S_{2ij}=\bordermatrix{%
&1  &\quad &i&\quad&j&\quad\cr
1 &0&\cdots&0&\cdots&0&\cdots\cr
\quad&\vdots&\cdots&\vdots&\cdots&\vdots&\cdots\cr
i &0&\cdots&a_{ij}&\cdots&-a_{ij}&\cdots\cr
&\vdots&\cdots&\vdots&\cdots&\vdots&\cdots\cr
j&0&\cdots&-a_{ij}&\cdots&a_{ij}&\cdots\cr
&\vdots&\cdots&\vdots&\cdots&\vdots&\cdots\cr
}_{N\times N}\nonumber.
\end{eqnarray}}%
Correspondingly, for $1\leqslant j\leqslant  i\leqslant N$, we get
{\setlength\abovedisplayskip{2pt}
\setlength\belowdisplayskip{2pt}
  \begin{eqnarray}S_{2ij}^{\rm T}S_{2ij}=\bordermatrix{%
&1  &\quad &j&\quad&i&\quad\cr
1 &0&\cdots&0&\cdots&0&\cdots\cr
\quad&\vdots&\cdots&\vdots&\cdots&\vdots&\cdots\cr
j &0&\cdots&a_{ij}&\cdots&-a_{ij}&\cdots\cr
&\vdots&\cdots&\vdots&\cdots&\vdots&\cdots\cr
i&0&\cdots&-a_{ij}&\cdots&a_{ij}&\cdots\cr
&\vdots&\cdots&\vdots&\cdots&\vdots&\cdots\cr
}_{N\times N}.
\nonumber
\end{eqnarray}}%
By Assumption 6, we have $a_{ij}=a_{ji}$, $1\leqslant  i,j\leqslant  N$, which together with the above equation gives
$\sum\limits_{i,j=1}^{N}\left({S_{2ij}^{\rm T}}S_{2ij}\right)=2\mathcal{L}$. $\hfill\square$

\section{Proof of theorems}
\emph{Proof of Theorem \ref{theorem1}:}
It can be seen that the (II) of Theorem \ref{theorem1} is a constructive method of the (I) of Theorem \ref{theorem1}.  Therefore,
if we can prove the (II) of Theorem \ref{theorem1}, then the (I) of Theorem \ref{theorem1} follows immediately. Our first goal is to show
the (II) of Theorem \ref{theorem1} holds.

Denote
$\hat{e}_i(t)$ $=x_i(t)-\hat{x}_i(t)$,
$\delta_i(t)=\hat{x}_{i0}(t)-x_0(t)$ and
$\delta(t)=(\delta^{\rm {T}}_1(t),\delta^{\rm {T}}_2(t),\ldots,$ $\delta^{\rm {T}}_N(t))^{\rm {T}}$.
By $(\ref{b1})$ and $(\ref{b3})$, we  get
{\setlength\abovedisplayskip{4pt}
\setlength\belowdisplayskip{4pt}
  \begin{eqnarray}\label{z0}
\dot{\hat{e}}_i(t)
 =&&\hspace{-.7cm}A_ix_i(t)+B_i u_i(t)-A_i\hat{x}_i(t)-B_i u_i(t)-H_i\left(y_i(t)-C_i\hat{x}_i(t)\right)\nonumber\\
 =&&\hspace{-.7cm}\left(A_i-H_i C_i\right)\hat{e}_i(t), \quad i=1,\ldots,N.
\end{eqnarray}}%
Based on Assumption \ref{A2}, we  choose $H_i$ such that $A_i-H_i C_i$ is Hurwitz, $i=1,\ldots,N$. By Lemma \ref{L2} and the above equation, we have
{\setlength\abovedisplayskip{4pt}
\setlength\belowdisplayskip{4pt}
  \begin{eqnarray}\label{z0001}
  \lim_{t\rightarrow \infty}\mathbb{E}\left[\left\| x_i(t)-\hat{x}_i(t)\right\|^2\right]=0,\quad i=1,\ldots,N.
  \end{eqnarray}}%
  In the following, we will  estimate
$\limsup\limits_{t\rightarrow \infty}\mathbb{E}\Big[\| \hat{x}_{i0}(t)-x_0(t)\|^2\Big], i=1,\ldots,N$.

Choose $G_{11}=\ldots=G_{1N}=G_1$ and $G_{21}=\ldots=G_{2N}=G_2$, where $G_1=k_1PC_0^{\rm T}\left(I_p +C_0PC_0^{\rm T}\right)^{-1}$ and $G_2=k_2PC_0^{\rm T}(I_p+C_0PC_0^{\rm T})^{-1}$.
In view of $(\ref{b2})$ and $(\ref{b4})$, we get
{\setlength\abovedisplayskip{4pt}
\setlength\belowdisplayskip{4pt}
  \begin{eqnarray}\label{klpz0001}
\textrm{d}\delta(t)
=&&\hspace{-.7cm}\left(I_N\otimes A_0-\mathcal{L}\otimes G_1C_0-F\otimes G_2C_0\right)\delta(t)\textrm{d}t+M_1\textrm{d}w_{1ij}(t)\nonumber\\
&&\hspace{-.7cm}+M_2\textrm{d}w_{1i0}(t)+M_3(t)\textrm{d}w_{2ij}(t)+M_4(t)\textrm{d}w_{2i0}(t),
\end{eqnarray}}%
 where $M_1=\sum\limits_{i,j=1}^{N}\left[S_{1ij}\otimes G_1
  \right]\Upsilon_{j}$,
   $M_2=\sum\limits_{i=1}^{N}\big[\bar{S}_{1i}\otimes G_2 \big]\Upsilon_{0}$,
  $M_3(t)=\sum\limits_{i,j=1}^{N}\sigma_{2ij}[S_{2ij}\otimes G_1C_0]\delta(t)$,
 $M_4(t)=-\sum\limits_{i=1}^{N}\sigma_{2i0}[\bar{S}_{2i}
\otimes G_2C_0]\delta(t)$.
Here, $\Upsilon_{j}= (\Upsilon^{\rm {T}}_{1j},\Upsilon^{\rm {T}}_{2j},\ldots,\Upsilon^{\rm {T}}_{Nj})^{\rm {T}}$ and $\Upsilon_{0}= (\Upsilon^{\rm {T}}_{10},\Upsilon^{\rm {T}}_{20},\ldots,\Upsilon^{\rm {T}}_{N0})^{\rm {T}}$; $S_{1ij}=[n_{kl}]_{N\times N}$ is an $N\times N$ matrix satisfying when $k=i$ and $l=i$, $n_{kl}=a_{ij}$,  and when $k$ and $l$ take other values, $n_{kl}=0$;
  $\bar{S}_{1i}=[\bar{n}_{kl}]_{N\times N}$ is an $N\times N$ matrix satisfying when  $k=i$ and $l=i$, $\bar{n}_{kl}=a_{i0}$, and when $k$ and $l$ take other values,
  $\bar{n}_{kl}=0$;
  $S_{2ij}=[s_{kl}]_{N\times N}$ is an $N\times N$  matrix  satisfying when $k=i$ and $l=i$, $s_{kl}=-a_{ij}$, when $k=i$ and $l=j$, $s_{kl}=a_{ij}$, and when $k$ and $l$ take other values, $s_{kl}=0$;
   $\bar{S}_{2i}=[\bar{s}_{kl}]_{N\times N}$ is an $N\times N$ matrix satisfying when $k=i$ and $l=i$, $\bar{s}_{kl}=a_{i0}$, and
   when $k$ and $l$ take other values, $\bar{s}_{kl}=0$.

From Assumption \ref{A6}, we know that $\mathcal{L}+F$ is a real symmetric matrix and all of its eigenvalues are positive.
 Hence, there exists a unitary matrix $\Phi$ such that $\Phi^{\rm T}(\mathcal{L}+F)\Phi=diag\big(\lambda_1(\mathcal{L}+F),\ldots,$ $\lambda_N(\mathcal{L}+F)\big)=:\Lambda$.

Denote $\bar{\delta}(t)=\left(\Phi^{-1}\otimes I_n\right)\delta(t)$. By  $(\ref{klpz0001})$, we have
{\setlength\abovedisplayskip{4pt}
\setlength\belowdisplayskip{4pt}
\begin{eqnarray}
\textrm{d}\bar{\delta}(t)=&&\hspace{-.7cm}\big(I_N\otimes A_0-\Phi^{\rm T}\mathcal{L}\Phi\otimes G_1C_0-\Phi^{\rm T}F\Phi\otimes G_2C_0\big)\bar{\delta}(t)\textrm{d}t\nonumber\\
&&\hspace{-.7cm}+M_5\textrm{d}w_{1ij}(t)+M_6\textrm{d}w_{1i0}(t)+M_7(t)\textrm{d}w_{2ij}(t)+M_8(t)\textrm{d}w_{2i0}(t),\nonumber
\end{eqnarray}}%
where $M_5=\sum\limits_{i,j=1}^{N}[\Phi^{\rm T}S_{1ij}\otimes G_1]\Upsilon_{j}$,
$M_6=\sum\limits_{i=1}^{N}[\Phi^{\rm T}\bar{S}_{1i}\otimes G_2]\Upsilon_0$,
$M_7(t)=\sum\limits_{i,j=1}^{N}\sigma_{2ij}[\Phi^{\rm T}S_{2ij}\Phi\otimes G_1C_0]\bar{\delta}(t)$,
$M_8(t)=-\sum\limits_{i=1}^{N}\sigma_{2i0}[\Phi^{\rm T}\bar{S}_{2i}\Phi\otimes G_2C_0]\bar{\delta}(t)$.

Choose the Lyapunov function $V(t)=\bar{\delta}^{\rm T}(t)\big(I_N\otimes P^{-1}\big)\bar{\delta}(t)$. Combining the above equation and It\^{o}'s formula, we obtain
{\setlength\abovedisplayskip{1pt}
\setlength\belowdisplayskip{1pt}
 \begin{eqnarray}\label{Z3}
\textrm{d}V(t)
=&&\hspace{-.7cm}2\bar{\delta}^{\rm T}(t)\left(I_N\otimes P^{-1}\right)\Big[I_N\otimes A_0-\Phi^{\rm T}\mathcal{L}\Phi\otimes G_1C_0-\Phi^{\rm T}F\Phi\otimes G_2C_0\Big]\bar{\delta}(t)\textrm{d}t+M_9(t)\textrm{d}t\nonumber\\
&&\hspace{-.7cm}+2\bar{\delta}^{\rm T}(t)\left(I_N\otimes P^{-1}\right)M_5\textrm{d}w_{1ij}(t)+2\bar{\delta}^{\rm T}(t)\left(I_N\otimes P^{-1}\right)M_6\textrm{d}w_{1i0}(t)\nonumber\\
&&\hspace{-.7cm}+2\bar{\delta}^{\rm T}(t)\left(I_N\otimes P^{-1}\right)M_7(t)\textrm{d}w_{2ij}(t)+2\bar{\delta}^{\rm T}(t)\left(I_N\otimes P^{-1}\right)M_8(t)\textrm{d}w_{2i0}(t),
\end{eqnarray}}
where $M_9(t)=\sum\limits_{i,j=1}^{N}
\Upsilon^{\rm T}_j\left[\left(\Phi^{\rm T}{S_{1ij}^{\rm T}}S_{1ij}\Phi\right)\otimes G_1^{\rm T}P^{-1}G_1\right]\Upsilon_j
+\sum\limits_{i=1}^{N}\Upsilon^{\rm T}_0\big[\left(\Phi^{\rm T}\bar{S}^{\rm T}_{1i}\bar{S}_{1i}\Phi\right)
\otimes G_2^{\rm T}P^{-1}G_2\big]\Upsilon_0
+\sum\limits_{i,j=1}^{N}\sigma_{ij}^2\bar{\delta}^{\rm T}(t)
\Big[\Big(\Phi^{\rm T}{S_{2ij}^{\rm T}}S_{2ij}\Phi\Big)\otimes C_0^{\rm T}G_1^{\rm T}P^{-1}G_1C_0\Big]\bar{\delta}(t)+\sum\limits_{i=1}^{N}\sigma_{i0}^2\bar{\delta}^{\rm T}(t)\big[\left(\Phi^{\rm T}
\bar{S}^{\rm T}_{2i}\bar{S}_{2i}\Phi\right)C_0^{\rm T}G_2^{\rm T}P^{-1} G_2C_0\big]\bar{\delta}(t)$.\\
Noting that $\sum\limits_{i=1}^{N}\left({S_{1ij}^{\rm T}}S_{1ij}\right)\leqslant  I_N,1\leqslant j\leqslant N$, $\sum\limits_{i=1}^{N}
\left(\bar{S}^{\rm T}_{1i}\bar{S}_{1i}\right)\leqslant I_N$,
$\sum\limits_{i,j=1}^{N}\left({S_{2ij}^{\rm T}}S_{2ij}\right)=2\mathcal{L}$ and $\sum\limits_{i=1}^{N}\left(\bar{S}^{\rm T}_{2i}\bar{S}_{2i}\right)=F\leqslant 2F$,
by the definition of  $M_9(t)$, we get
{\setlength\abovedisplayskip{1pt}
\setlength\belowdisplayskip{1pt}
\begin{eqnarray}
M_9(t)
\leqslant&&\hspace{-.7cm}\sum\limits_{j=1}^{N}
\Upsilon^{\rm T}_j\left[\sum\limits_{i=1}^{N}\left(\Phi^{\rm T}{S_{1ij}^{\rm T}}S_{1ij}\Phi\right)\otimes G_1^{\rm T}P^{-1}G_1\right]\Upsilon_j+\Upsilon^{\rm T}_0\left[\sum\limits_{i=1}^{N}\left(\Phi^{\rm T}\bar{S}^{\rm T}_{1i}\bar{S}_{1i}\Phi\right)\otimes G_2^{\rm T}P^{-1}G_2\right]\Upsilon_0\nonumber\\
&&\hspace{-.7cm}+ \sigma^2
\bar{\delta}^{\rm T}(t)\Bigg[\left(\sum\limits_{i,j=1}^{N}\Phi^{\rm T}{S_{2ij}^{\rm T}}S_{2ij}\Phi\right)
\otimes C_0^{\rm T}G_1^{\rm T}P^{-1}G_1C_0\Bigg]\bar{\delta}(t)\nonumber\\
&&\hspace{-.7cm}+ \sigma^2\bar{\delta}^{\rm T}(t)\Bigg[\left(\sum\limits_{i=1}^{N}\Phi^{\rm T}
\bar{S}^{\rm T}_{2i}\bar{S}_{2i}\Phi\right)\otimes C_0^{\rm T}G_2^{\rm T}P^{-1}G_2C_0\Bigg]\bar{\delta}(t)\nonumber\\
\leqslant&&\hspace{-.6cm}\sum\limits_{j=1}^{N}
\Upsilon^{\rm T}_j\left[I_N\otimes G_1^{\rm T}P^{-1}G_1\right]\Upsilon_j+\Upsilon^{\rm T}_0\left[I_N\otimes G_2^{\rm T}P^{-1}G_2\right]\Upsilon_0\nonumber\\
&&\hspace{-.6cm}+2 \sigma^2\bar{\delta}^{\rm T}(t)\left(\Phi^{\rm T}\mathcal{L}\Phi\otimes C_0^{\rm T}G_1^{\rm T}P^{-1}G_1C_0\right)\bar{\delta}(t)\nonumber\\
&&\hspace{-.6cm}+2 \sigma^2\bar{\delta}^{\rm T}(t)\left(\Phi^{\rm T}F\Phi\otimes C_0^{\rm T}G_2^{\rm T}P^{-1}G_2C_0\right)\bar{\delta}(t),\nonumber
\end{eqnarray}}%
where $\sigma^2=\max\bigg\{\max\limits_{1\leqslant i,j\leqslant N} \sigma_{ij}^2,\max\limits_{1\leqslant i\leqslant N} \sigma_{i0}^2\bigg\}$.
In view of $(\ref{Z3})$  and the above inequality, we have
{\setlength\abovedisplayskip{1pt}
\setlength\belowdisplayskip{1pt}
 \begin{eqnarray}
\textrm{d}V(t)
\leqslant&&\hspace{-.7cm}2\bar{\delta}^{\rm T}(t)\left(I_N\otimes P^{-1}\right)\big[I_N\otimes A_0-\Phi^{\rm T}\mathcal{L}\Phi\otimes G_1C_0-\Phi^{\rm T}F\Phi\otimes G_2C_0\big]\bar{\delta}(t)\textrm{d}t\nonumber\\
&&\hspace{-.7cm}+2 \sigma^2\bar{\delta}^{\rm T}(t)\big[\Phi^{\rm T}\mathcal{L}\Phi\otimes C_0^{\rm T}G_1^{\rm T}P^{-1}G_1C_0+\Phi^{\rm T}F\Phi\otimes C_0^{\rm T}G_2^{\rm T}P^{-1}G_2C_0\big]\bar{\delta}(t)\textrm{d}t\nonumber\\
&&\hspace{-.7cm}+\varpi_1\textrm{d}t+2\bar{\delta}^{\rm T}(t)\left(I_N\otimes P^{-1}\right)M_5\textrm{d}w_{1ij}(t)+2\bar{\delta}^{\rm T}(t)\left(I_N\otimes P^{-1}\right)M_6\textrm{d}w_{1i0}(t)\nonumber\\
&&\hspace{-.7cm}+2\bar{\delta}^{\rm T}(t)\left(I_N\otimes P^{-1}\right)M_7(t)\textrm{d}w_{2ij}(t)
+2\bar{\delta}^{\rm T}(t)\left(I_N\otimes P^{-1}\right)M_8(t)\textrm{d}w_{2i0}(t),\nonumber
\end{eqnarray}}%
where $\varpi_1= \sum\limits_{j=1}^{N}
\Upsilon^{\rm T}_j\big[I_N\otimes G_1^{\rm T}P^{-1}G_1\big]\Upsilon_j+\Upsilon^{\rm T}_0\big[I_N\otimes G_2^{\rm T}P^{-1}G_2\big]\Upsilon_0$.\\
Denote $\tilde{\delta}(t)=\left(I_N\otimes P^{-1}\right)\bar{\delta}(t)$, which together with  the above inequality gives
{\setlength\abovedisplayskip{1pt}
 \setlength\belowdisplayskip{1pt}
\begin{eqnarray}
\textrm{d}V(t)
\leqslant&&\hspace{-.7cm}-2\tilde{\delta}^{\rm T}(t)\left[\Phi^{\rm T}\mathcal{L}\Phi\otimes G_1C_0P+\Phi^{\rm T}F\Phi\otimes G_2C_0P\right]\tilde{\delta}(t)\textrm{d}t+2 \sigma^2\tilde{\delta}^{\rm T}(t)\big[\Phi^{\rm T}\mathcal{L}\Phi\otimes PC_0^{\rm T}G_1^{\rm T}P^{-1}\nonumber\\
&&\hspace{-.7cm}\times G_1C_0P+\Phi^{\rm T}F\Phi\otimes PC_0^{\rm T}G_2^{\rm T}P^{-1}G_2C_0P\big]\tilde{\delta}(t)\textrm{d}t+\varpi_1\textrm{d}t\nonumber\\
&&\hspace{-.7cm}+2\tilde{\delta}^{\rm T}(t)\left(I_N\otimes A_0\right)\left(I_N\otimes P\right)\tilde{\delta}(t)\textrm{d}t+2\tilde{\delta}^{\rm T}(t)M_5\textrm{d}w_{1ij}(t)+2\tilde{\delta}^{\rm T}(t)M_6\textrm{d}w_{1i0}(t)\nonumber\\
&&\hspace{-.7cm}+2\tilde{\delta}^{\rm T}(t)M_{10}(t)\textrm{d}w_{2ij}(t)+2\tilde{\delta}^{\rm T}(t)M_{11}(t)\textrm{d}w_{2i0}(t),\nonumber
\end{eqnarray}}%
where $ M_{10}(t)=\sum\limits_{i,j=1}^{N}\sigma_{2ij}[\Phi^{\rm T}S_{2ij}\Phi\otimes G_1C_0P]\tilde{\delta}(t)$ and
$ M_{11}(t)=-\sum\limits_{i=1}^{N}\sigma_{2i0}[\Phi^{\rm T}\bar{S}_{2i}\Phi\otimes G_2C_0P] \tilde{\delta}(t)$.\\
Denote $W(t)=e^{\gamma t}V(t)$, $0<\gamma<\frac{1}{\|P\|}$. By the above inequality and applying  It\^{o}'s formula to $W(t)$, we  get
 {\setlength\abovedisplayskip{1pt}
  \setlength\belowdisplayskip{1pt}
   \begin{eqnarray}
\textrm{d}W (t)
=&&\hspace{-.7cm}\gamma e^{\gamma t}V(t)\textrm{d}t+e^{\gamma t}\textrm{d}V(t)\nonumber\\
\leqslant&&\hspace{-.7cm}\gamma e^{\gamma t}\left(\tilde{\delta}^{\rm T}(t)\left(I_N\otimes P\right)\tilde{\delta}(t)\right)\textrm{d}t+ e^{\gamma t}\big[\tilde{\delta}^{\rm T}(t)\big(I_N\otimes \big(A_0P+PA_0^{\rm T}\big)\big)\tilde{\delta}(t)\big]\textrm{d}t\nonumber\\
&&\hspace{-.7cm}-2e^{\gamma t}\tilde{\delta}^{\rm T}(t)\big[\Phi^{\rm T}\mathcal{L}\Phi\otimes G_1C_0P+\Phi^{\rm T}F\Phi\otimes G_2C_0P\big]\tilde{\delta}(t)\textrm{d}t\nonumber\\
&&\hspace{-.7cm}+2\sigma^2e^{\gamma t}\tilde{\delta}^{\rm T}(t)\big[\Phi^{\rm T}\mathcal{L}\Phi
\otimes PC_0^{\rm T}G_1^{\rm T}P^{-1}G_1C_0P+\Phi^{\rm T}F\Phi\otimes PC_0^{\rm T}G_2^{\rm T}P^{-1}G_2C_0P\big]\tilde{\delta}(t)\textrm{d}t\nonumber\\
&&\hspace{-.7cm}+e^{\gamma t}\varpi_1\textrm{d}t
+2e^{\gamma t}\tilde{\delta}^{\rm T}(t)M_5\textrm{d}w_{1ij}(t)+2e^{\gamma t}\tilde{\delta}^{\rm T}(t)M_6\textrm{d}w_{1i0}(t)\nonumber\\
&&\hspace{-.7cm}+2e^{\gamma t}\tilde{\delta}^{\rm T}(t)M_{10}(t)\textrm{d}w_{2ij}(t)+2e^{\gamma t}\tilde{\delta}^{\rm T}(t)M_{11}(t)\textrm{d}w_{2i0}(t),
\nonumber\\
=&&\hspace{-.7cm}e^{\gamma t}\tilde{\delta}^{\rm T}(t)\Psi(\gamma)
\tilde{\delta}(t)\textrm{d}t+ e^{\gamma t}\varpi_1\textrm{d}t+2e^{\gamma t}\tilde{\delta}^{\rm T}(t)M_5\textrm{d}w_{1ij}(t)+2e^{\gamma t}\tilde{\delta}^{\rm T}(t)M_6\textrm{d}w_{1i0}(t)\nonumber\\
&&\hspace{-.7cm}+2e^{\gamma t}\tilde{\delta}^{\rm T}(t)M_{10}(t)\textrm{d}w_{2ij}(t)+2e^{\gamma t}\tilde{\delta}^{\rm T}(t)M_{11}(t)\textrm{d}w_{2i0}(t),\nonumber
\end{eqnarray}}%
where
$\Psi\left(\gamma\right)=\gamma\left(I_N\otimes P\right)+I_N\otimes \left(A_0P+PA_0^{\rm T}\right)-2 \Phi^{\rm T}\mathcal{L}\Phi\otimes G_1C_0P-2\Phi^{\rm T}F\Phi\otimes G_2C_0P
+2 \sigma^2 \Phi^{\rm T}\mathcal{L}\Phi\otimes PC_0^{\rm T}G_1^{\rm T}P^{-1}G_1C_0P+2 \sigma^2\Phi^{\rm T}F\Phi\otimes PC_0^{\rm T}G_2^{\rm T}P^{-1}G_2C_0P$.\\
 Integrating both sides of the above inequality from $0$ to $t$ and taking the
mathematical expectation, we obtain
{\setlength\abovedisplayskip{1pt}
\setlength\belowdisplayskip{1pt}
\begin{eqnarray}\label{a2Z9}
\mathbb{E}[W (t)]
\leqslant&&\hspace{-.7cm}\mathbb{E}[W (0)]+ \int_0^te^{\gamma s}\varpi_1\textrm{d}s+\mathbb{E}\left[\int_0^te^{\gamma s}\left[\tilde{\delta}^{\rm T}(s)\Psi\left(\gamma\right)\tilde{\delta}(s)\right]\textrm{d}s\right].
\end{eqnarray}}%
In the following, we proceed to prove that the matrix $\Psi(\gamma)<0$.
Noting that $G_1=k_1PC_0^{\rm T}(I_p+C_0PC_0^{\rm T})^{-1}$ and $G_2=k_2PC_0^{\rm T}(I_p+C_0PC_0^{\rm T})^{-1}$, we have
{\setlength\abovedisplayskip{1pt}
\setlength\belowdisplayskip{1pt}
\begin{eqnarray}\label{afv2Z9}
\Psi\left(\gamma\right)
\leqslant&&\hspace{-.7cm}\gamma\left(I_N\otimes P\right)+I_N\otimes \left(A_0P+PA_0^{\rm T}\right)-2 k_1\left(\Phi^{\rm T}\mathcal{L}\Phi\right)\otimes PC_0^{\rm T}\left(I_p+C_0PC_0^{\rm T}\right)^{-1}C_0P\nonumber\\
&&\hspace{-.7cm}
-2 k_2\left(\Phi^{\rm T}F\Phi\right)\otimes PC_0^{\rm T}\left(I_p+C_0PC_0^{\rm T}\right)^{-1}C_0P+2 \sigma^2k_1^2 \left(\Phi^{\rm T}\mathcal{L}\Phi\right)\otimes PC_0^{\rm T}\left(I_p+C_0PC_0^{\rm T}\right)^{-1}C_0P\nonumber\\
&&\hspace{-.7cm}
+2 \sigma^2k_2^2 \left(\Phi^{\rm T}F\Phi\right)\otimes PC_0^{\rm T}\left(I_p+C_0PC_0^{\rm T}\right)^{-1}C_0P\nonumber\\
\leqslant&&\hspace{-.7cm}\gamma\left(I_N\otimes P\right)+I_N\otimes \left(A_0P+PA_0^{\rm T}\right)-\frac{2\alpha}{\lambda_1(\mathcal{L}+F)}
\left(\Phi^{\rm T}(\mathcal{L}+F)\Phi\right)\otimes PC_0^{\rm T}\left(I_p+C_0PC_0^{\rm T}\right)^{-1}C_0P\nonumber\\
=&&\hspace{-.7cm}\gamma\left(I_N\otimes P\right)+I_N\otimes \left(A_0P+PA_0^{\rm T}\right)-\frac{2\alpha\Lambda}{\lambda_1(\mathcal{L}+F)}\otimes PC_0^{\rm T}\left(I_p+C_0PC_0^{\rm T}\right)^{-1}C_0P.
\end{eqnarray}}%
If $\mathbb{R}\mathbbm{e}\left(\lambda_n\left(A_0\right)\right)\geqslant 0$ holds, by Assumption \ref{A3} and choosing $\alpha>\lambda_0^{u}(A_0 ) $, we know that  Lemma \ref{L1} holds. By the generalized algebraic Riccati equation $(\ref{a0})$, we get
{\setlength\abovedisplayskip{1pt}
\setlength\belowdisplayskip{1pt}
\begin{eqnarray}\label{afv2Z10}
&&\hspace{-.7cm}\gamma P+A_0P+PA_0^{\rm T}-\frac{2\alpha\lambda_i(\mathcal{L}+F)}{\lambda_1(\mathcal{L}+F)} PC_0^{\rm T}
(I_p+C_0PC_0^{\rm T})^{-1}C_0P\nonumber\\
=&&\hspace{-.7cm}\gamma P+2\alpha PC_0^{\rm T}\left(I_p+C_0PC_0^{\rm T}\right)^{-1}C_0P-I_n -\frac{2\alpha\lambda_i(\mathcal{L}+F)}{\lambda_1(\mathcal{L}+F)} PC_0^{\rm T}\left(I_p+C_0PC_0^{\rm T}\right)^{-1}C_0P\nonumber\\
\leqslant&&\hspace{-.7cm}\gamma P+2\alpha PC_0^{\rm T}\left(I_p+C_0PC_0^{\rm T}\right)^{-1}C_0P-I_n-2\alpha PC_0^{\rm T}\left(I_p+C_0PC_0^{\rm T}\right)^{-1}C_0P\nonumber\\
=&&\hspace{-.7cm}\gamma P-I_n<0,\quad i=1,\ldots,N.
\end{eqnarray}}%
If $\mathbb{R}\mathbbm{e}\left(\lambda_n\left(A_0\right)\right)< 0$ holds, by Assumption \ref{A3} and choosing $\alpha=0$, we know that  Lemma \ref{L1} holds. By the Lyapunov equation $(\ref{a0})$, we get
{\setlength\abovedisplayskip{1pt}
\setlength\belowdisplayskip{1pt}
\begin{eqnarray}
\gamma P+A_0P+PA_0^{\rm T}-\frac{2\alpha\lambda_i(\mathcal{L}+F)}{\lambda_1(\mathcal{L}+F)} PC_0^{\rm T}
(I_p+C_0PC_0^{\rm T})^{-1}C_0P
=\gamma P-I_n<0,\quad i=1,\ldots,N.\nonumber
\end{eqnarray}}
Therefore, by  $(\ref{a2Z9})$$-$$(\ref{afv2Z10})$ and  the above equation,  we  obtain
{\setlength\abovedisplayskip{1pt}
 \setlength\belowdisplayskip{1pt}
 \begin{eqnarray}
\mathbb{E}[W(t)]
\leqslant \mathbb{E}[W(0)]+\varpi_1 \int_0^te^{\gamma s}\textrm{d}s,\nonumber
\end{eqnarray}}%
which together with the definition of $W(t)$ gives
{\setlength\abovedisplayskip{1pt}
\setlength\belowdisplayskip{1pt}
 \begin{eqnarray}
 \mathbb{E}[V(t)]\leqslant e^{-\gamma t}\mathbb{E}[V (0)]+
\varpi_1\int_0^te^{\gamma (s-t)}\textrm{d}s.\nonumber
\end{eqnarray}}%
Noting that
$\lambda_{\max}^{-1}(P)\mathbb{E}\left[\left\|\bar{\delta}(t)\right\|^2\right]\leqslant \mathbb{E}[V(t)]$,  by the above inequality, we get
{\setlength\abovedisplayskip{1pt}
\setlength\belowdisplayskip{1pt}
\begin{eqnarray}\label{Z14}
 \mathbb{E}\left[\left\|\bar{\delta}(t)\right\|^2\right]
 \leqslant\lambda_{\max}(P)e^{-\gamma t}\mathbb{E}[V(0)]+\frac{\lambda_{\max}(P)\varpi_1}{\gamma}\left(1-e^{-\gamma t}\right),
\end{eqnarray}}%
which implies
{\setlength\abovedisplayskip{1pt}
\setlength\belowdisplayskip{1pt}
 \begin{eqnarray}
\limsup\limits_{t\rightarrow \infty}\mathbb{E}\left[\left\|\bar{\delta}(t)\right\|^2\right]\leqslant\frac{\lambda_{\max}(P)\varpi_1}{\gamma}.\nonumber
\end{eqnarray}}%
Then, noting that $\bar{\delta}(t)=\left(\Phi^{-1}\otimes I_n\right)\delta(t)$,  we have
{\setlength\abovedisplayskip{1pt}
 \setlength\belowdisplayskip{1pt}
 \begin{eqnarray}
\limsup\limits_{t\rightarrow \infty}\mathbb{E}\left[\left\|\delta(t)\right\|^2\right]=\limsup\limits_{t\rightarrow \infty}\mathbb{E}\left[\left\|\bar{\delta}(t)\right\|^2\right]\leqslant\frac{\lambda_{\max}(P)\varpi_1}{\gamma}.\nonumber
\end{eqnarray}}%
This together with the definition of $\delta(t)$ gives
{\setlength\abovedisplayskip{1pt}
\setlength\belowdisplayskip{1pt}
\begin{eqnarray}
\limsup\limits_{t\rightarrow \infty}\mathbb{E}\Big[\| \hat{x}_{i0}(t)-x_0(t)\|^2\Big]\leqslant\frac{\lambda_{\max}(P)\varpi_1}{\gamma}.\nonumber
\end{eqnarray}}%
Then, we proceed to estimate $\limsup\limits_{t\rightarrow \infty}\mathbb{E}\left[\left\|y_i(t)- y_0(t)\right\|^2\right]$ for $i=1,\ldots,N$.

Denote $\Delta_i(t)=\hat{x}_i(t)-\Pi_ix_0(t),i=1,\ldots,N$.
Noting that $K_{2i}=\Gamma_i-K_{1i}\Pi_i,i=1,\ldots,N$, by $(\ref{b2})$, $(\ref{b3})$ and Assumption \ref{A4}, we have
 {\setlength\abovedisplayskip{1pt}
  \setlength\belowdisplayskip{1pt}
   \begin{eqnarray}
\dot{\Delta}_i(t)
=&&\hspace{-.7cm} {A_i}\hat{x}_{i}(t)+B_{i}u_{i}(t)+H_i\left(y_i(t)-\hat{y}_{i}(t)\right)-\Pi_i A_0x_0(t)\nonumber\\
=&&\hspace{-.7cm} {A_i}\hat{x}_{i}(t)+B_{i}u_{i}(t)+H_i\left(y_i(t)-\hat{y}_{i}(t)\right)-\left(A_i\Pi_i+B_i\Gamma_i\right)x_0(t)\nonumber\\
=&&\hspace{-.7cm} {A_i}\Delta_i(t)+B_{i}\left(K_{1i}\hat{x}_i(t)+K_{2i}\hat{x}_{i0}(t)\right)
-B_i\Gamma_ix_0(t)+H_i\left(y_i(t)-\hat{y}_{i}(t)\right)\nonumber\\
=&&\hspace{-.7cm} {A_i}\Delta_i(t)+B_{i}K_{1i}\hat{x}_i(t)+B_{i}
\left(\Gamma_i-K_{1i}\Pi_i\right)\hat{x}_{i0}(t)-B_i\Gamma_ix_0(t)+H_i C_i\left(x_i(t)-\hat{x}_{i}(t)\right)\nonumber\\
=&&\hspace{-.7cm} {A_i}\Delta_i(t)+B_{i}K_{1i}\left(\hat{x}_i(t)-\Pi_ix_0(t)\right)+B_{i}K_{1i}\Pi_i
(x_0(t) -\hat{x}_{i0}(t))+B_i\Gamma_i\big(\hat{x}_{i0}(t)-x_0(t)\big)\nonumber\\
&&\hspace{-.7cm}+H_i C_i\left(x_i(t)-\hat{x}_{i}(t)\right)\nonumber\\
=&&\hspace{-.7cm} \left({A_i}+B_{i}K_{1i}\right)\Delta_i(t) +
\left(B_i\Gamma_i-B_{i}K_{1i}\Pi_i\right)\left(\hat{x}_{i0}(t)-x_0(t)\right)+H_i C_i\left(x_i(t)-\hat{x}_{i}(t)\right)\nonumber\\
=&&\hspace{-.7cm} \left({A_i}+B_{i}K_{1i}\right)\Delta_i(t) +B_i K_{2i}\left(\hat{x}_{i0}(t)-x_0(t)\right)+H_i C_i\left(x_i(t)-\hat{x}_{i}(t)\right).\nonumber
 \end{eqnarray}}%
 Integrating both sides of  the above equation from 0 to $t$ yields
 {\setlength\abovedisplayskip{1pt}
 \setlength\belowdisplayskip{1pt}
 \begin{eqnarray}
\Delta_i(t)
=&&\hspace{-.7cm}e^{(A_i+B_iK_{1i})t}\Delta_i(0)+\int_0^te^{(A_i+B_iK_{1i})(t-s)}B_i K_{2i}
\left(\hat{x}_{i0}(s)-x_0(s)\right)\textrm{d}s\nonumber\\
&&\hspace{-.7cm}+\int_0^te^{(A_i+B_iK_{1i})(t-s)}H_i C_i\left(x_i(s)-\hat{x}_{i}(s)\right)\textrm{d}s,\nonumber
 \end{eqnarray}}%
which implies
 {\setlength\abovedisplayskip{1pt}
   \setlength\belowdisplayskip{1pt}
    \begin{eqnarray}\label{Zz20}
\mathbb{E}\left[\left\|\Delta_i(t)\right\|^2\right]
\leqslant&&\hspace{-.7cm}3\mathbb{E}\left[\left\|e^{(A_i+B_iK_{1i})t}\Delta_i(0)\right\|^2\right]
 +3\mathbb{E}\bigg[\Big
\|\int_0^te^{(A_i+B_iK_{1i})(t-s)} B_i K_{2i}(\hat{x}_{i0}(s)-x_0(s))\textrm{d}s\Big\|^2\bigg]\nonumber\\
&&\hspace{-.7cm} +3\mathbb{E}\Bigg[\bigg\|\int_0^te^{(A_i+B_iK_{1i})(t-s)} H_i C_i\left(x_i(s)-\hat{x}_{i}(s)\right)\textrm{d}s\bigg\|^2\Bigg].
 \end{eqnarray}}%
Noting that
$\left(E\left[\left\|\int_{0}^{t}X(\tau)\textrm{d}\tau\right\|^2\right]\right)^{1/2}\leqslant\int_{0}^{t}\left(E\left[
\|X(\tau)\|^2\right]\right)^{1/2}\textrm{d}\tau$, we have
{\setlength\abovedisplayskip{1pt}
\setlength\belowdisplayskip{1pt}
\begin{eqnarray}\label{Zz21}
&&\hspace{-.7cm}\left(\mathbb{E}\left[\left\|\int_0^te^{(A_i+B_iK_{1i})(t-s)}B_i K_{2i}\left(\hat{x}_{i0}(s)-x_0(s)\right)\textrm{d}s
\right\|^2\right]\right)^{\frac{1}{2}}\nonumber\\
\leqslant&&\hspace{-.7cm}\int_0^t\left(\mathbb{E}\left[\left\|e^{(A_i+B_iK_{1i})(t-s)}B_i K_{2i}\left(\hat{x}_{i0}(s)-x_0(s)\right)
\right\|^2\right]\right)^{\frac{1}{2}}\textrm{d}s\nonumber\\
\leqslant&&\hspace{-.7cm}\int_0^t\Bigg(\left\|e^{(A_i+B_iK_{1i})(t-s)}\right\|^2\left\|B_i K_{2i}\right\|^2\mathbb{E}
\left[\left\|\hat{x}_{i0}(s)-x_0(s)\right\|^2\right]\Bigg)^{\frac{1}{2}}\textrm{d}s.
 \end{eqnarray}}%
 By Assumption \ref{A1}, we choose $K_{1i}$ such that ${A_i}+B_{i}K_{1i}$ is Hurwitz matrix.
 As ${A_i}+B_{i}K_{1i}$ is Hurwitz,  there exist positive constants $\rho_1>0$ and $\rho_2>0$ such that
 {\setlength\abovedisplayskip{1pt}
 \setlength\belowdisplayskip{1pt}
 \begin{eqnarray}\label{Zz22}
 \left\|e^{(A_i+B_iK_{1i})t}\right\|\leqslant\rho_1 e^{-\rho_2 t}, \quad i=1,\ldots,N.
  \end{eqnarray}}%
  Since $\limsup\limits_{t\rightarrow \infty}\mathbb{E}\left[\left\|\delta(t)\right\|^2\right]\leqslant\frac{\lambda_{\max}(P)\varpi_1}{\gamma}$,  for any given constant $\varepsilon>0$, there exists a positive constant $T_1>0$ such that for any $t>T_1$
  {\setlength\abovedisplayskip{1pt}
  \setlength\belowdisplayskip{1pt}
  \begin{eqnarray}
  \mathbb{E}\left[\left\|\hat{x}_{i0}(t)-x_0(t)\right\|^2\right]\leqslant\frac{\lambda_{\max}(P)\varpi_1}{\gamma}
  +\varepsilon, \quad i=1,\ldots,N,\nonumber
 \end{eqnarray}}%
which together with $(\ref{Zz21})$ and $(\ref{Zz22})$ gives
  {\setlength\abovedisplayskip{1pt}
  \setlength\belowdisplayskip{1pt}
  \begin{eqnarray}
&&\hspace{-.7cm}\left(\mathbb{E}\left[\left\|\int_0^te^{(A_i+B_iK_{1i})(t-s)}B_i K_{2i}
\left(\hat{x}_{i0}(s)-x_0(s)\right)\textrm{d}s\right\|^2\right]\right)^{\frac{1}{2}}\nonumber\\
\leqslant&&\hspace{-.7cm}\int_0^t\left(\mathbb{E}\left[\left\|e^{(A_i+B_iK_{1i})(t-s)}B_i K_{2i}\left(\hat{x}_{i0}(s)-x_0(s)\right)
\right\|^2\right]\right)^{\frac{1}{2}}\textrm{d}s\nonumber\\
\leqslant&&\hspace{-.7cm}\left\|B_i K_{2i}\right\|\int_0^{T_1}\rho_1 e^{-\rho_2 (t-s)}\left(\mathbb{E}\left[\left\|\hat{x}_{i0}(s)-x_0(s)\right\|^2\right]\right)^{\frac{1}{2}}\textrm{d}s\nonumber\\
&&\hspace{-.7cm}+\left(\frac{\lambda_{\max}(P)\varpi_1}{\gamma}+\varepsilon\right)\left\|B_i K_{2i}\right\|\int_{T_1}^t\rho_1 e^{-\rho_2 (t-s)}\textrm{d}s\nonumber\\
=&&\hspace{-.7cm}\left\|B_i K_{2i}\right\|e^{-\rho_2 t}\int_0^{T_1}\rho_1 e^{\rho_2s}\left(\mathbb{E}\left[\left\|\hat{x}_{i0}(s)-x_0(s)\right\|^2\right]\right)^{\frac{1}{2}}\textrm{d}s\nonumber\\
&&\hspace{-.7cm}+\left(\frac{\lambda_{\max}(P)\varpi_1}{\gamma}+\varepsilon\right)\frac{\rho_1
\left\|B_i K_{2i}\right\|}{\rho_2} \left(1-e^{\rho_2 (T_1-t)}\right).\nonumber
 \end{eqnarray}}%
 Taking the limit on  both sides of the above inequality, we get
 {\setlength\abovedisplayskip{1pt}
  \setlength\belowdisplayskip{1pt}
    \begin{eqnarray}
\limsup\limits_{t\rightarrow \infty}\Bigg(\mathbb{E}\Bigg[\bigg\|\int_0^te^{(A_i+B_iK_{1i})(t-s)} B_i K_{2i}\left(\hat{x}_{i0}(s)-x_0(s)\right)
\textrm{d}s\bigg\|^2\Bigg]\Bigg)^{\frac{1}{2}}
\leqslant\left(\frac{\lambda_{\max}(P)\varpi_1}{\gamma}+\varepsilon\right)
\frac{\rho_1\|B_i K_{2i}\|}{\rho_2}.\nonumber
 \end{eqnarray}}%
Then, by  the arbitrariness of $\varepsilon$, we obtain
{\setlength\abovedisplayskip{1pt}
\setlength\belowdisplayskip{1pt}
\begin{eqnarray}
\limsup\limits_{t\rightarrow \infty}\Bigg(\mathbb{E}\Bigg[\bigg\|\int_0^te^{(A_i+B_iK_{1i})(t-s)} B_i K_{2i}\left(\hat{x}_{i0}(s)-x_0(s)\right)
\textrm{d}s\bigg\|^2\Bigg]\Bigg)^{\frac{1}{2}}
\leqslant\frac{\rho_1\lambda_{\max}(P)\varpi_1}{\gamma\rho_2}\left\|B_iK_{2i}\right\|,\nonumber
 \end{eqnarray}}%
which gives
 {\setlength\abovedisplayskip{1pt}
 \setlength\belowdisplayskip{1pt}
 \begin{eqnarray}\label{Zz25}
\limsup\limits_{t\rightarrow \infty}\mathbb{E}\Bigg[\bigg\|\int_0^te^{(A_i+B_iK_{1i})(t-s)}B_iK_{2i}
\left(\hat{x}_{i0}(s)-x_0(s)\right)\textrm{d}s\bigg\|^2\Bigg]
\leqslant\frac{\rho_1^2\lambda^2_{\max}(P)\varpi_1^2}{\rho_2^2\gamma^2}\left\|B_iK_{2i}\right\|^2.
 \end{eqnarray}}%
Noting  that ${\lim\limits_{t\rightarrow \infty}}\mathbb{E}\left[\left\|x_i(t)-\hat{x}_{i}(t)\right\|^2\right]=0$, similar to the above inequality, we get
{\setlength\abovedisplayskip{1pt}
 \setlength\belowdisplayskip{1pt}
  \begin{eqnarray}\label{yZz25}
\limsup\limits_{t\rightarrow \infty}\mathbb{E}\left[\left\|\int_0^te^{(A_i+B_iK_{1i})(t-s)}H_iC_i
(x_i(s)-\hat{x}_{i}(s))\textrm{d}s\right\|^2\right]=0.\nonumber
  \end{eqnarray}}%
 By $(\ref{Zz20})$, $(\ref{Zz22})$$-$$(\ref{Zz25})$ and the above equation, we obtain
  {\setlength\abovedisplayskip{1pt}
  \setlength\belowdisplayskip{1pt}
  \begin{eqnarray}\label{yZz26}
\limsup\limits_{t\rightarrow \infty}\mathbb{E}\left[\left\|\hat{x}_i(t)-\Pi_ix_0(t)\right\|^2\right]
\leqslant\frac{3\rho_1^2\lambda^2_{\max}(P)
\varpi_1^2}{\rho_2^2\gamma^2}\left\|B_i K_{2i}\right\|^2.
  \end{eqnarray}}%
  By  the $C_r$ inequality, $(\ref{b2})$$-$$(\ref{b3})$ and  $(\ref{a1})$, we have
 {\setlength\abovedisplayskip{1pt}
 \setlength\belowdisplayskip{1pt}
 \begin{eqnarray}\label{Z2kl4}
\mathbb{E}\left[\left\|y_i(t)-y_0(t)\right\|^2\right]
\leqslant&&\hspace{-.7cm}2\mathbb{E}\left[\left\| y_i(t)-\hat{y}_i(t)\right\|^2\right]+2\mathbb{E}\left[\left\|\hat{y}_i(t)-y_0(t)\right\|^2\right]\nonumber\\
\leqslant&&\hspace{-.7cm}2\left\|C_i\right\|^2 \mathbb{E}\left[\left\| x_i(t)-\hat{x}_i(t)\right\|^2\right]
+2\left\| C_i\right\|^2\mathbb{E}\left[\left\|\hat{x}_i(t)-\Pi_ix_0(t)\right\|^2\right]\nonumber,
\end{eqnarray}}%
which together with $(\ref{z0001})$ and  $(\ref{yZz26})$ gives
{\setlength\abovedisplayskip{1pt}
\setlength\belowdisplayskip{1pt}
\begin{eqnarray}\label{Z2l4}
\limsup\limits_{t\rightarrow \infty}\mathbb{E}\left[\left\|y_i(t)-y_0(t)\right\|^2\right]
\leqslant\frac{6\rho_1^2\lambda^2_{\max}(P)\varpi_1^2}{\rho_2^2\gamma^2}\left\| C_i\right\|^2\left\|B_iK_{2i}\right\|^2,\quad i=1,\ldots,N.\nonumber
\end{eqnarray}}%
For any given constant $ \varepsilon> 0$, by the above inequality and $0<\gamma<\frac{1}{\|P\|}$,
we have
{\setlength\abovedisplayskip{1pt}
\setlength\belowdisplayskip{1pt}
\begin{eqnarray}
\limsup\limits_{t\rightarrow \infty}\mathbb{E}\left[\left\|y_i(t)-y_0(t)\right\|^2\right]
\leqslant\frac{6\rho_1^2\lambda^2_{\max}(P)\varpi_1^2}{\rho_2^2\left(\frac{1}{\|P\|}-\varepsilon\right)^2}\left\| C_i\right\|^2\left\|B_iK_{2i}\right\|^2,\quad i=1,\ldots,N,\nonumber
\end{eqnarray}}%
which together with  the arbitrariness of $\varepsilon$ gives
{\setlength\abovedisplayskip{1pt}
\setlength\belowdisplayskip{1pt}
\begin{eqnarray}\label{Z24}
\limsup\limits_{t\rightarrow \infty}\mathbb{E}\left[\left\|y_i(t)-y_0(t)\right\|^2\right]
\leqslant\frac{6\rho_1^2\lambda^2_{\max}(P)\varpi_1^2\|P\|^2}{\rho_2^2}\left\| C_i\right\|^2\left\|B_iK_{2i}\right\|^2,\quad i=1,\ldots,N.
\end{eqnarray}}%
Therefore, the leader-following HMASs $(\ref{b2})$$-$$(\ref{b1})$ achieve mean square bounded output tracking and  the proof of (II) of Theorem \ref{theorem1} is completed. Finally,
the (I) of Theorem \ref{theorem1}  follows immediately from what we have proved before. $\hfill\square$
\vskip 0.2cm

\emph{Proof of Theorem  \ref{theorem3}}:
In particular, if there are no additive measurement noises, by the definition of $\varpi_1$,  we get
$\varpi_1=0$. Noting that $\varpi_1=0$,
 by  $(\ref{Z2l4})$, we get
{\setlength\abovedisplayskip{1pt}
\setlength\belowdisplayskip{1pt}
\begin{eqnarray}
\limsup\limits_{t\rightarrow \infty}\mathbb{E}\left[\left\| y_i(t)-y_0(t)\right\|^2\right]=0, \quad i=1,\ldots,N,\nonumber
 \end{eqnarray}}which together with
$\mathbb{E}\left[\left\| y_i(t)-y_0(t)\right\|^2\right]\geqslant0$
gives
{\setlength\abovedisplayskip{1pt}
\setlength\belowdisplayskip{1pt}
\begin{eqnarray}
\lim\limits_{t\rightarrow \infty}\mathbb{E}\left[\left\| y_i(t)-y_0(t)\right\|^2\right]=0, \quad i=1, \ldots, N. \nonumber\end{eqnarray}}%
Therefore, the leader-following HMASs $(\ref{b2})$$-$$(\ref{b1})$ achieve mean square output tracking. Similar to the proof of Theorem \ref{theorem1}$-$(I), we know that
there exist admissible observation  and   cooperative control strategies  such that  the HMASs $(\ref{b2})$$-$$(\ref{b1})$ achieve mean square output tracking.

In the following, we will   estimate
the mean square output tracking time.\\
By $(\ref{Z24})$ and the definitions of $\hat{e}_i(t)$ and $\Delta_i(t)$, we  get
{\setlength\abovedisplayskip{1pt}
\setlength\belowdisplayskip{1pt}
\begin{eqnarray}\label{qqsZ24}
\mathbb{E}\left[\left\|y_i(t)-y_0(t)\right\|^2\right]
\leqslant 2\left\|C_i\right\|^2 \mathbb{E}\left[\left\|\hat{e}_i(t)\right\|^2\right]+2\left\| C_i\right\|^2\mathbb{E}\left[\left\|\Delta_i(t)\right\|^2\right].
\end{eqnarray}}%
By $(\ref{z0})$, we have
 {\setlength\abovedisplayskip{1pt}
 \setlength\belowdisplayskip{1pt}
  \begin{eqnarray}\label{qwvz0}
\hat{e}_i(t)=e^{(A_i-H_i C_i)t}\hat{e}_i(0), \quad i=1, \ldots, N.
\end{eqnarray}}%
Substituting $(\ref{Zz20})$ and $(\ref{qwvz0})$ into $(\ref{qqsZ24})$ leads to
{\setlength\abovedisplayskip{1pt}
\setlength\belowdisplayskip{1pt}
\begin{eqnarray}\label{qqsZ25}
&&\hspace{-.7cm}\mathbb{E}\left[\left\|y_i(t)-y_0(t)\right\|^2\right]\nonumber\\
\leqslant
&&\hspace{-.7cm}6\left\|C_i\right\|^2\mathbb{E}\Bigg[\bigg\|\int_0^te^{(A_i+B_iK_{1i})(t-s)}B_i K_{2i}(\hat{x}_{i0}(s)-x_0(s))\textrm{d}s\bigg\|^2\Bigg]\nonumber\\
&&\hspace{-.7cm}
+6\left\|C_i\right\|^2\mathbb{E}\Bigg[\bigg\|\int_0^te^{(A_i+B_iK_{1i})(t-s)} H_i C_i\left(x_i(s)-\hat{x}_{i}(s)\right)\textrm{d}s\bigg\|^2\Bigg]\nonumber\\
&&\hspace{-.7cm}
+6\left\|C_i\right\|^2\mathbb{E}\Bigg[\bigg\|e^{(A_i+B_iK_{1i})t}\Delta_i(0)\bigg\|^2\Bigg]
+2\left\|C_i\right\|^2 \mathbb{E}\left[\left\|e^{(A_i-H_i C_i)t}\hat{e}_i(0)\right\|^2\right].
\end{eqnarray}}%
Noting that $\varpi_1=0$, by $(\ref{Z14})$ and the definition of $\bar{\delta}(t)$, we get
{\setlength\abovedisplayskip{1pt}
\setlength\belowdisplayskip{1pt}
\begin{eqnarray}
\mathbb{E}
\left[\left\|\hat{x}_{i0}(t)-x_0(t)\right\|^2\right]\leqslant\lambda_{\max}(P)e^{-\gamma t}\mathbb{E}\left[V (0)\right].\nonumber
 \end{eqnarray}}%
Noting that
$\left(E\left[\left\|\int_{0}^{t}X(\tau)\textrm{d}\tau\right\|^2\right]\right)^{1/2}\leqslant\int_{0}^{t}\left(E\left[
\|X(\tau)\|^2\right]\right)^{1/2}\textrm{d}\tau$, by the above inequality,
we have
{\setlength\abovedisplayskip{4pt}
\setlength\belowdisplayskip{4pt}
 \begin{eqnarray}\label{qqsZ26}
&&\hspace{-.7cm}\left(\mathbb{E}\left[\left\|\int_0^te^{(A_i+B_iK_{1i})(t-s)}B_i K_{2i}
\left(\hat{x}_{i0}(s)-x_0(s)\right)\textrm{d}s\right\|^2\right]\right)^{\frac{1}{2}}\nonumber\\
\leqslant&&\hspace{-.7cm}\int_0^t\Bigg(\left\|e^{(A_i+B_iK_{1i})(t-s)}\right\|^2\left\|B_i K_{2i}\right\|^2\lambda_{\max}(P)e^{-\gamma s}\mathbb{E}\left[V (0)\right]\Bigg)^{\frac{1}{2}}\textrm{d}s\nonumber\\
\leqslant&&\hspace{-.7cm}\left(\lambda_{\max}(P)\mathbb{E}[V (0)]\right)^{\frac{1}{2}}\left\|B_i K_{2i}\right\|\int_0^t\left\|e^{(A_i+B_iK_{1i})(t-s)}\right
\|e^{-\frac{\gamma}{2}s}
\textrm{d}s.
 \end{eqnarray}}%
Noting that
$\left(E\left[\left\|\int_{0}^{t}X(\tau)\textrm{d}\tau\right\|^2\right]\right)^{1/2}\leqslant\int_{0}^{t}\left(E\left[
\|X(\tau)\|^2\right]\right)^{1/2}\textrm{d}\tau$, by $(\ref{qwvz0})$,
we obtain
 {\setlength\abovedisplayskip{1pt}
 \setlength\belowdisplayskip{1pt}
  \begin{eqnarray}
&&\hspace{-.7cm}\left(\mathbb{E}\left[\left\|\int_0^te^{(A_i+B_iK_{1i})(t-s)}
H_i C_i\left(x_i(s)-\hat{x}_{i}(s)\right)\textrm{d}s\right\|^2\right]\right)^{\frac{1}{2}}\nonumber\\
\leqslant&&\hspace{-.7cm}\int_0^t\Bigg(\left\|e^{(A_i+B_iK_{1i})(t-s)}\right\|^2\left\|H_i C_i\right\|^2\mathbb{E}\left[\left\|e^{(A_i-H_i C_i)s}\hat{e}_i(0)\right\|^2\right]\Bigg)^{\frac{1}{2}}\textrm{d}s\nonumber\\
=&&\hspace{-.7cm}\left\|H_i C_i\right\|\int_0^t\left\|e^{(A_i+B_iK_{1i})(t-s)}\right\|\left(\mathbb{E}\left[\left\|e^{(A_i-H_i C_i)s}\hat{e}_i(0)\right\|^2\right]\right)^\frac{1}{2}\textrm{d}s.\nonumber
 \end{eqnarray}}%
Substituting $(\ref{qqsZ26})$ and the above inequality into $(\ref{qqsZ25})$ leads to
 {\setlength\abovedisplayskip{1pt}
  \setlength\belowdisplayskip{1pt}
    \begin{eqnarray}\label{qqsZ28}
&&\hspace{-.7cm}\mathbb{E}\left[\left\|y_i(t)-y_0(t)\right\|^2\right]\nonumber\\
\leqslant&&\hspace{-.7cm}2\left\|C_i\right\|^2 \mathbb{E}\left[\left\|\hat{e}_i(0)\right\|^2\right]\left\|e^{(A_i-H_i C_i)t}\right\|^2
+6\left\|C_i\right\|^2\mathbb{E}\left[\left\|\Delta_i(0)\right\|^2\right]\left\|e^{(A_i+B_iK_{1i})t}\right\|^2\nonumber\\
&&\hspace{-.7cm}+6\left\|C_i\right\|^2\left\|B_i K_{2i}\right\|^2
\lambda_{\max}(P)\mathbb{E}\left[V (0)\right]\left(\int_0^t\left\|e^{(A_i+B_iK_{1i})(t-s)}\right\|e^{-\frac{\gamma}{2} s}
\textrm{d}s\right)^2\nonumber\\
&&\hspace{-.7cm}+6\left\|C_i\right\|^2\mathbb{E}\Big[\|\hat{e}_i(0)\|^2\Big]
\left\|H_i C_i\right\|^2\left(\int_0^t\left\|e^{(A_i+B_iK_{1i})(t-s)}\right\|
\left\|e^{(A_i-H_i C_i)s}\right\|\textrm{d}s\right)^2.
\end{eqnarray}}%
As  $A_i-H_i C_i$ is Hurwitz,  there exist positive constants $\rho_3>0$ and $\rho_4>0$ such that
 {\setlength\abovedisplayskip{1pt}
  \setlength\belowdisplayskip{1pt}
    \begin{eqnarray}\label{dsZz22}
\left \|e^{(A_i-H_i C_i)t}\right\|\leqslant\rho_3 e^{-\rho_4 t},\quad  i=1,\ldots,N.
  \end{eqnarray}}%
 By the definitions of $V(t)$, $\bar{\delta}(t)$, $\delta(t)$ and $\Phi$, we get
  {\setlength\abovedisplayskip{1pt}
  \setlength\belowdisplayskip{1pt}
  \begin{eqnarray}\label{dsdfp23}
\mathbb{E}\left[V(0)\right]
\leqslant &&\hspace{-.7cm}\lambda_{\max}(I_N\otimes P^{-1})\mathbb{E}\left[\|\bar{\delta}(0)\|^2\right]\nonumber\\
\leqslant&&\hspace{-.7cm}\lambda^{-1}_{\min}(I_N\otimes P)\|\Phi^{-1}\otimes I_n\|\mathbb{E}\left[\sum\limits_{i=1}^{N}\left\|\hat{x}_{i0}(0)-x_0(0)\right\|\right]\nonumber\\
=&&\hspace{-.7cm}\lambda^{-1}_{\min}(P)\mathbb{E}\left[\sum\limits_{i=1}^{N}\left\|\hat{x}_{i0}(0)-x_0(0)\right\|\right].
  \end{eqnarray}}%
By $(\ref{Zz22})$, $(\ref{qqsZ28})-(\ref{dsdfp23})$ and the definitions of  $\hat{e}_i(t)$ and $\Delta_i(t)$, we have
{\setlength\abovedisplayskip{1pt}
 \setlength\belowdisplayskip{1pt}
  \begin{eqnarray}\label{qqsZ29}
\mathbb{E}\left[\left\|y_i(t)-y_0(t)\right\|^2\right]
\leqslant&&\hspace{-.7cm}2\rho_3^2 e^{-2\rho_4 t}\left\|C_i\right\|^2 \mathbb{E}\left[\left\|\hat{e}_i(0)\right\|^2\right]
+6\rho_1^2e^{-2\rho_2 t}\left\|C_i\right\|^2\mathbb{E}\left[\left\|\Delta_i(0)\right\|^2\right]\nonumber\\
&&\hspace{-.7cm}+6\left\|C_i\right\|^2\left\|B_i K_{2i}\right\|^2\lambda_{\max}(P)\mathbb{E}\left[V (0)\right]\left(\int_0^t\rho_1 e^{-\rho_2 (t-s)}e^{-\frac{\gamma}{2} s}
\textrm{d}s\right)^2\nonumber\\
&&\hspace{-.7cm}+6\left\|C_i\right\|^2\left\|H_i C_i\right\|^2\mathbb{E}\left[\left\|\hat{e}_i(0)\right\|^2\right]\left(\int_0^t\rho_1\rho_3 e^{-\rho_2 (t-s)-\rho_4 s}\textrm{d}s\right)^2\nonumber\\
\leqslant&&\hspace{-.7cm}2\rho_3^2 e^{-2\rho_4 t}\left\|C_i\right\|^2 \mathbb{E}\left[\left\|\hat{e}_i(0)\right\|^2\right]
+6\rho_1^2e^{-2\rho_2 t}\left\|C_i\right\|^2\mathbb{E}\left[\left\|\Delta_i(0)\right\|^2\right]\nonumber\\
&&\hspace{-.7cm}+6\rho_1^2\left\|C_i\right\|^2\left\|B_i K_{2i}\right\|^2\lambda_{\max}(P)\mathbb{E}\left[V (0)\right]
\left(te^{-\min\{\rho_2,\frac{\gamma}{2}\} t}\right)^2\nonumber\\
&&\hspace{-.7cm}+6\rho_1^2\rho_3^2 \left\|C_i\right\|^2\left\|H_i C_i\right\|^2\mathbb{E}\left[\left\|\hat{e}_i(0)\right\|^2\right]\left(te^{-\min\{\rho_2,\rho_4\} t}\right)^2.\nonumber
\end{eqnarray}}%
For $0<\tau_1 <\min\{\rho_2,\frac{\gamma}{2}\}$ and $0<\tau_2 <\min\{\rho_2,\rho_4\}$, there exist positive constants $T_2(\tau_1)=\frac{2}{\tau_1^2}$ and $T_3(\tau_2)=\frac{2}{\tau_2^2}$ such that for any $t>T_2(\tau_1)$ and  $t>T_3(\tau_2)$, we have $e^{\tau_1 t}>\frac{\tau_1^2t^2}{2}>t$ and $e^{\tau_2 t}>\frac{\tau_2^2t^2}{2}>t$.\\
By the above inequality, we get
{\setlength\abovedisplayskip{1pt}
 \setlength\belowdisplayskip{1pt}
  \begin{eqnarray}\label{qqsZv210}
\mathbb{E}\left[\left\|y_i(t)-y_0(t)\right\|^2\right]
\leqslant&&\hspace{-.7cm}2\rho_3^2 e^{-2\rho_4 t}\left\|C_i\right\|^2 \mathbb{E}\left[\left\|\hat{e}_i(0)\right\|^2\right]
+6\rho_1^2e^{-2\rho_2 t}\left\|C_i\right\|^2\mathbb{E}\left[\left\|\Delta_i(0)\right\|^2\right]\nonumber\\
&&\hspace{-.7cm}+6\rho_1^2\left\|C_i\right\|^2\left\|B_i K_{2i}\right\|^2\lambda_{\max}(P)\mathbb{E}\left[V (0)\right]
e^{-2\big(\min\{\rho_2,\frac{\gamma}{2}\}-\tau_1\big)t}\nonumber\\
&&\hspace{-.7cm}+6\rho_1^2\rho_3^2 \left\|C_i\right\|^2\left\|H_i C_i\right\|^2\mathbb{E}\left[\left\|\hat{e}_i(0)\right\|^2\right]e^{-2\big(\min\{\rho_2,\rho_4\}-\tau_2\big)t}\nonumber\\
\leqslant&&\hspace{-.7cm}2\rho_3^2 e^{-2\rho_4 t}\left\|C_i\right\|^2 \mathbb{E}\left[\left\|x_i(0)-\hat{x}_i(0)\right\|^2\right]\nonumber\\
&&\hspace{-.7cm}+6\rho_1^2e^{-2\rho_2 t}\left\|C_i\right\|^2\mathbb{E}\Big[\big\|\hat{x}_i(0)-\Pi_ix_0(0)\big\|^2\Big]\nonumber\\
&&\hspace{-.7cm}+\frac{6\rho_1^2\left\|C_i\right\|^2\left\|B_i K_{2i}\right\|^2\lambda_{\max}(P)}{\lambda_{\min}(P)}\mathbb{E}\left[\left
\|x_i(0)-\hat{x}_i(0)\right\|^2\right] e^{-2\big(\min\{\rho_2,\frac{\gamma}{2}\}-\tau_1\big)t}\nonumber\\
&&\hspace{-.7cm}
+6\rho_1^2\rho_3^2 \left\|C_i\right\|^2\left\|H_i C_i\right\|^2\mathbb{E}\left[\left\|x_i(0)-\hat{x}_i(0)\right\|^2\right]e^{-2\big(\min\{\rho_2,\rho_4\}-\tau_2\big)t}\nonumber\\
\leqslant&&\hspace{-.7cm}\Bigg\{2\rho_3^2\left\|C_i\right\|^2 \mathbb{E}\left[\left\|x_i(0)-\hat{x}_i(0)\right\|^2\right]
+6\rho_1^2\left\|C_i\right\|^2\mathbb{E}\Big[\big\|\hat{x}_i(0)-\Pi_ix_0(0)\big\|^2\Big]\nonumber\\
&&\hspace{-.7cm}+\frac{6\rho_1^2\left\|C_i\right\|^2\left\|B_i K_{2i}\right\|^2\lambda_{\max}(P)}{\lambda_{\min}(P)}\mathbb{E}\left[\left
\|x_i(0)-\hat{x}_i(0)\right\|^2\right]+6\rho_1^2\rho_3^2 \left\|C_i\right\|^2\nonumber\\
&&\hspace{-.7cm}
\times\left\|H_i C_i\right\|^2
\mathbb{E}\left[\left\|x_i(0)-\hat{x}_i(0)\right\|^2\right]\Bigg\} e^{-2\min\left\{\rho_2-\tau_1,\rho_2-\tau_2,\rho_4-\tau_2,\frac{\gamma}{2}-\tau_1\right\}t}\nonumber\\
\leqslant&&\hspace{-.7cm}\varpi_2e^{-2\min\left\{\rho_2-\tau_1,\rho_2-\tau_2,\rho_4-\tau_2,\frac{\gamma}{2}-\tau_1\right\}t},\quad i=1,\ldots,N.\nonumber
\end{eqnarray}}%
Therefore, for $0<\varepsilon<\varpi_2$, there exists a positive constant $T_4(\varepsilon,\gamma,\tau_1,\tau_2)=
\max\Bigg\{\frac{2}{\tau_1^2},\frac{2}{\tau_2^2},$\\$\frac{\ln\left(\frac{\varpi_2}{\varepsilon}\right)}{2\min\left\{\rho_2-\tau_1,\rho_2-\tau_2,
\rho_4-\tau_2,\frac{\gamma}{2}-\tau_1\right\}}\Bigg\}$ such that for any $t>T_4(\varepsilon,\gamma,\tau_1,\tau_2)$, we get
{\setlength\abovedisplayskip{1pt}
\setlength\belowdisplayskip{1pt}
\begin{eqnarray}\label{dsqqsZ29}
\mathbb{E}\left[\left\|y_i(t)-y_0(t)\right\|^2\right]
\leqslant\varpi_2e^{-2\min\left\{\rho_2-\tau_1,\rho_2-\tau_2,\rho_4-\tau_2,\frac{\gamma}{2}-\tau_1\right\}t}
\leqslant \varepsilon.\nonumber
\end{eqnarray}}%
By  $0<\gamma<\frac{1}{\|P\|}$, $0<\tau_1 <\min\{\rho_2,\frac{\gamma}{2}\}$ and $0<\tau_2 <\min\{\rho_2,\rho_4\}$, we have
{\setlength\abovedisplayskip{1pt}
\setlength\belowdisplayskip{1pt}
\begin{eqnarray}\label{dsqqsZ29}
\inf\limits_{\gamma,\tau_1,\tau_2}T_4(\varepsilon,\gamma,\tau_1,\tau_2)
=\max\left\{\frac{2}{\min\{\rho_2^2,\rho_4^2,\frac{1}{4\|P\|^2}\}},
\frac{\ln\left(\frac{\varpi_2}{\varepsilon}\right)}{2\min\left\{\rho_2,
\rho_4,\frac{1}{2\|P\|}\right\}}\right\}.
\end{eqnarray}}%
For $\varepsilon\geqslant\varpi_2$, we  know  that for $\forall t\geqslant0$,
{\setlength\abovedisplayskip{1pt}
\setlength\belowdisplayskip{1pt}
 \begin{eqnarray}
\mathbb{E}\left[\left\|y_i(t)-y_0(t)\right\|^2\right]
\leqslant\varpi_2e^{-2\min\left\{\rho_2-\tau_1,\rho_2-\tau_2,\rho_4-\tau_2,\frac{\gamma}{2}-\tau_1\right\}t}
\leqslant \varepsilon,\nonumber
\end{eqnarray}}%
which together with $(\ref{dsqqsZ29})$ gives
 {\setlength\abovedisplayskip{1pt}
  \setlength\belowdisplayskip{1pt}
   \begin{eqnarray}\label{lkvbjjuna1}
\left\{ \begin{array}{l}
t_{\varepsilon}\leqslant\max\left\{\frac{2}{\min\{\rho_2^2,\rho_4^2,\frac{1}{4\|P\|^2}\}},
\frac{\ln\left(\frac{\varpi_2}{\varepsilon}\right)}{2\min\left\{\rho_2,
\rho_4,\frac{1}{2\|P\|}\right\}}\right\},0<\varepsilon<\varpi_2,
 \\[0.5em]
t_{\varepsilon}=0, \quad \varepsilon\geqslant\varpi_2.
\end{array} \right.\nonumber
\end{eqnarray}}%
 $\hfill\square$

\vskip 0.2cm

\emph{Proof of Theorem \ref{theorem2}}:
Denote
$e_i(t)$ $=x_i(t)-\hat{x}_i(t)$ and
$\delta_i(t)=\hat{x}_{i0}(t)-x_0(t)$, $i=1,\ldots N$.
By $(\ref{b1})$ and $(\ref{b3})$, we  get
{\setlength\abovedisplayskip{4pt}
\setlength\belowdisplayskip{4pt}
  \begin{eqnarray}\label{ll0}
\dot{e}_i(t)
 =&&\hspace{-.7cm}a_ix_i(t)+a_i u_i(t)-a_i\hat{x}_i(t)-a_i u_i(t)-h_i\left(y_i(t)-c_i\hat{x}_i(t)\right)\nonumber\\
 =&&\hspace{-.7cm}\left(a_i-h_i c_i\right)e_i(t), \quad i=1,\ldots,N,\nonumber
\end{eqnarray}}%
which implies
{\setlength\abovedisplayskip{4pt}
\setlength\belowdisplayskip{4pt}
  \begin{eqnarray}\label{lpn1}
e_i(t)=e^{\left(a_i-h_i c_i\right)t}e_i(0),\quad i=1,\ldots N.
\end{eqnarray}}%
Choose $G_{21}=\ldots=G_{2N}=k$,
by $(\ref{b2})$, $(\ref{b4})$ and Assumption $\ref{Ak2}$, we get
{\setlength\abovedisplayskip{4pt}
\setlength\belowdisplayskip{4pt}
  \begin{eqnarray}
\textrm{d}\delta_i(t)
=\left(a_0-kc_0\right)\delta_i(t)\textrm{d}t+k\Upsilon_{i0}\textrm{d}w_{1i0}(t)-k\sigma_{i0}c_0\delta_i(t)\textrm{d}w_{2i0}(t),\nonumber
\end{eqnarray}}%
which together with the Theorem 3.1 of \cite{Mao33} gives
{\setlength\abovedisplayskip{4pt}
\setlength\belowdisplayskip{4pt}
  \begin{eqnarray}\label{vlpn2}
\delta_i(t)
=&&\hspace{-.7cm}e^{\left(a_0-kc_0-\frac{1}{2}k^2\sigma_{i0}^2c_0^2\right)t-k\sigma_{i0}c_0w_{2i0}(t)}\delta_i(0)+k\Upsilon_{i0}\int_0^te^{\left(a_0-kc_0
-\frac{1}{2}k^2\sigma_{i0}^2c_0^2\right)(t-s)}\textrm{d}w_{1i0}(s)\nonumber\\
&&\hspace{-.7cm}+k^2\sigma_{i0}\Upsilon_{i0}c_0
\int_0^te^{\left(a_0-kc_0-\frac{1}{2}k^2\sigma_{i0}^2c_0^2\right)(t-s)-k\sigma_{i0}c_0(w_{2i0}(t)-w_{2i0}(s))}\textrm{d}s.\nonumber
\end{eqnarray}}%
Taking the
mathematical expectation, by Assumption $\ref{Ak1}$, we obtain
{\setlength\abovedisplayskip{4pt}
\setlength\belowdisplayskip{4pt}
  \begin{eqnarray}\label{vlpn3}
\mathbb{E}[\delta_i(t)]
=&&\hspace{-.7cm}e^{\left(a_0-kc_0-\frac{1}{2}k^2\sigma_{i0}^2c_0^2\right)t}\mathbb{E}[e^{-k\sigma_{i0}c_0w_{2i0}(t)}]\mathbb{E}[\delta_i(0)]\nonumber\\
&&\hspace{-.7cm}+k^2\sigma_{i0}\Upsilon_{i0}c_0\int_0^te^{\big(a_0-kc_0-\frac{1}{2}k^2\sigma_{i0}^2c_0^2\big)(t-s)}
\mathbb{E}\Big[e^{-k\sigma_{i0}c_0(w_{2i0}(t)-w_{2i0}(s))}\Big]\textrm{d}s\nonumber\\
=&&\hspace{-.7cm}e^{\left(a_0-kc_0\right)t}\mathbb{E}[\delta_i(0)]+k^2\sigma_{i0}\Upsilon_{i0}c_0\int_0^te^{\left(a_0-kc_0\right)(t-s)}\textrm{d}s\nonumber\\
=&&\hspace{-.7cm}e^{\left(a_0-kc_0\right)t}\left[\mathbb{E}[\delta_i(0)] +\frac{k^2\sigma_{i0}\Upsilon_{i0}c_0}{a_0-kc_0}\right]-\frac{k^2\sigma_{i0}\Upsilon_{i0}c_0}{a_0-kc_0}.
\end{eqnarray}}%
Denote $\Delta_i(t)=\hat{x}_i(t)-\pi_ix_0(t),i=1,\ldots,N$.
Noting that $k_{2i}=\gamma_i-k_{1i}\pi_i,i=1,\ldots,N$, by $(\ref{b2})$ and $(\ref{b3})$, we have
{\setlength\abovedisplayskip{1pt}
  \setlength\belowdisplayskip{1pt}
   \begin{eqnarray}
\dot{\Delta}_i(t)
=&&\hspace{-.7cm} {a_i}\hat{x}_{i}(t)+b_{i}u_{i}(t)+h_i\left(y_i(t)-\hat{y}_{i}(t)\right)-\pi_i a_0x_0(t)\nonumber\\
=&&\hspace{-.7cm} {a_i}\hat{x}_{i}(t)+b_{i}u_{i}(t)+h_i\left(y_i(t)-\hat{y}_{i}(t)\right)-\left(a_i\pi_i+b_i\gamma_i\right)x_0(t)\nonumber\\
=&&\hspace{-.7cm} \left({a_i}+b_{i}k_{1i}\right)\Delta_i(t) +b_i k_{2i}\left(\hat{x}_{i0}(t)-x_0(t)\right)+h_i c_i\left(x_i(t)-\hat{x}_{i}(t)\right).\nonumber
 \end{eqnarray}}%
 Integrating both sides of  the above equation from 0 to $t$ and taking the
mathematical expectation, we obtain
 {\setlength\abovedisplayskip{1pt}
 \setlength\belowdisplayskip{1pt}
 \begin{eqnarray}\label{vlpn4}
\mathbb{E}[\Delta_i(t)]
=&&\hspace{-.7cm}e^{(a_i+b_i k_{1i})t}\mathbb{E}[\Delta_i(0)]+\int_0^te^{(a_i+b_ik_{1i})(t-s)}b_i k_{2i}
\mathbb{E}[\hat{x}_{i0}(s)-x_0(s)]\textrm{d}s\nonumber\\
&&\hspace{-.7cm}+\int_0^te^{(a_i+b_ik_{1i})(t-s)}h_i c_i\mathbb{E}[x_i(s)-\hat{x}_{i}(s)]\textrm{d}s.
 \end{eqnarray}}%
 By $(\ref{lpn1})$, we get
 {\setlength\abovedisplayskip{1pt}
 \setlength\belowdisplayskip{1pt}
 \begin{eqnarray}\label{vlpn5}
\int_0^te^{(a_i+b_ik_{1i})(t-s)}h_i c_i\mathbb{E}[x_i(s)-\hat{x}_{i}(s)]\textrm{d}s
=h_i c_i\int_0^te^{(a_i+b_ik_{1i})(t-s)}e^{\left(a_i-h_i c_i\right)s}\mathbb{E}[e_i(0)]\textrm{d}s.
 \end{eqnarray}}%
If $h_ic_i=-b_ik_{1i}$, by $(\ref{vlpn5})$, we have
{\setlength\abovedisplayskip{1pt}
 \setlength\belowdisplayskip{1pt}
 \begin{eqnarray}\label{vlpn6}
\int_0^te^{(a_i+b_ik_{1i})(t-s)}h_i c_i\mathbb{E}[x_i(s)-\hat{x}_{i}(s)]\textrm{d}s
=h_i c_ie^{(a_i+b_ik_{1i})t}t\mathbb{E}[e_i(0)].
 \end{eqnarray}}%
 If $h_ic_i\neq-b_ik_{1i}$, by $(\ref{vlpn5})$, we have
{\setlength\abovedisplayskip{1pt}
 \setlength\belowdisplayskip{1pt}
 \begin{eqnarray}\label{vlpn7}
\int_0^te^{(a_i+b_ik_{1i})(t-s)}h_i c_i\mathbb{E}[x_i(s)-\hat{x}_{i}(s)]\textrm{d}s
=\frac{1}{h_i c_i+b_i k_{1i}}\left[e^{(a_i+b_ik_{1i})t}-e^{(a_i-h_ic_i)t}\right]\mathbb{E}[e_i(0)].\nonumber
 \end{eqnarray}}%
 Combing $(\ref{vlpn5})$, $(\ref{vlpn6})$ and  the above equation, we get
 {\setlength\abovedisplayskip{1pt}
 \setlength\belowdisplayskip{1pt}
 \begin{eqnarray}\label{vlpn8}
\lim\limits_{t\rightarrow \infty}\int_0^te^{(a_i+b_ik_{1i})(t-s)}h_i c_i\mathbb{E}[x_i(s)-\hat{x}_{i}(s)]\textrm{d}s=0.
 \end{eqnarray}}%
 By $(\ref{vlpn3})$, we have
 {\setlength\abovedisplayskip{1pt}
 \setlength\belowdisplayskip{1pt}
\begin{eqnarray}\label{vlpn9}
&&\hspace{-.7cm}\int_0^te^{(a_i+b_ik_{1i})(t-s)}b_i k_{2i}
\mathbb{E}[\hat{x}_{i0}(s)-x_0(s)]\textrm{d}s\nonumber\\
=&&\hspace{-.7cm}\int_0^te^{(a_i+b_ik_{1i})(t-s)}b_i k_{2i}
\Big[e^{\left(a_0-kc_0\right)t}\left(\mathbb{E}[\delta_i(0)] +\frac{k^2\sigma_{i0}\Upsilon_{i0}c_0}{a_0-kc_0}\right)-\frac{k^2\sigma_{i0}\Upsilon_{i0}c_0}{a_0-kc_0}\Big]\textrm{d}s\nonumber\\
=&&\hspace{-.7cm}b_i k_{2i}\left(\mathbb{E}[\delta_i(0)] +\frac{k^2\sigma_{i0}\Upsilon_{i0}c_0}{a_0-kc_0}\right)\int_0^te^{(a_i+b_ik_{1i})(t-s)+(a_0-kc_0)t}\textrm{d}s-b_i k_{2i}\frac{k^2\sigma_{i0}\Upsilon_{i0}c_0}{a_0-kc_0}\int_0^te^{(a_i+b_ik_{1i})(t-s)}\textrm{d}s\nonumber\\
=&&\hspace{-.7cm}b_i k_{2i}\left(\mathbb{E}[\delta_i(0)] +\frac{k^2\sigma_{i0}\Upsilon_{i0}c_0}{a_0-kc_0}\right)\frac{e^{(a_i+b_ik_{1i}+a_0-kc_0)t}-e^{(a_0-kc_0)t}}{a_i+b_ik_{1i}}-b_i k_{2i}\frac{k^2\sigma_{i0}\Upsilon_{i0}c_0}{a_0-kc_0}\frac{e^{(a_i+b_ik_{1i})t}-1}{a_i+b_ik_{1i}},\nonumber
 \end{eqnarray}}%
which implies
 {\setlength\abovedisplayskip{1pt}
 \setlength\belowdisplayskip{1pt}
 \begin{eqnarray}\label{vlpn10}
\lim\limits_{t\rightarrow \infty}\int_0^te^{(a_i+b_ik_{1i})(t-s)}b_i k_{2i}
\mathbb{E}[\hat{x}_{i0}(s)-x_0(s)]\textrm{d}s
=\frac{b_i k_{2i}k^2\sigma_{i0}\Upsilon_{i0}c_0}{(a_0-kc_0)(a_i+b_ik_{1i})}.\nonumber
 \end{eqnarray}}%
 By $(\ref{vlpn4})$, $(\ref{vlpn8})$ and the above equation, we have
 {\setlength\abovedisplayskip{1pt}
 \setlength\belowdisplayskip{1pt}
 \begin{eqnarray}\label{vlpn11}
\lim\limits_{t\rightarrow \infty}\mathbb{E}[\Delta_i(t)]
=\frac{b_i k_{2i}k^2\sigma_{i0}\Upsilon_{i0}c_0}{(a_0-kc_0)(a_i+b_ik_{1i})}.\nonumber
 \end{eqnarray}}%
 By  $(\ref{b2})$$-$$(\ref{b3})$ and  $(\ref{a1})$, we get
 {\setlength\abovedisplayskip{1pt}
 \setlength\belowdisplayskip{1pt}
 \begin{eqnarray}\label{vlpn12}
\mathbb{E}\left[y_i(t)-y_0(t)\right]
=&&\hspace{-.7cm}\mathbb{E}\left[ y_i(t)-\hat{y}_i(t)\right]+\mathbb{E}\left[\hat{y}_i(t)-y_0(t)\right]\nonumber\\
=&&\hspace{-.7cm}c_i\mathbb{E}\left[ x_i(t)-\hat{x}_i(t)\right]+c_i\mathbb{E}\left[\hat{x}_i(t)-\pi_ix_0(t)\right]\nonumber\\
=&&\hspace{-.7cm}c_i\mathbb{E}\left[ e_i(t)\right]+c_i\mathbb{E}\left[\Delta_i(t)\right],\quad i=1,\ldots N,\nonumber
\end{eqnarray}}%
 which together with $(\ref{lpn1})$ gives
 {\setlength\abovedisplayskip{1pt}
 \setlength\belowdisplayskip{1pt}
 \begin{eqnarray}\label{vlpn13}
\lim\limits_{t\rightarrow \infty}\mathbb{E}\left[y_i(t)-y_0(t)\right]
=\frac{c_i b_i k_{2i}k^2\sigma_{i0}\Upsilon_{i0}c_0}{(a_0-kc_0)(a_i+b_ik_{1i})}.\nonumber
 \end{eqnarray}}%
By the above equation, we get
{\setlength\abovedisplayskip{1pt}
 \setlength\belowdisplayskip{1pt}
 \begin{eqnarray}\label{vlpn14}
\liminf\limits_{t\rightarrow \infty}\mathbb{E}\left[\left|y_i(t)-y_0(t)\right|\right]
\geqslant&&\hspace{-.7cm}\lim\limits_{t\rightarrow \infty}\left|\mathbb{E}\left[y_i(t)-y_0(t)\right]\right|
=\frac{|c_i b_i k_{2i}k^2\sigma_{i0}\Upsilon_{i0}c_0|}{(a_0-kc_0)(a_i+b_ik_{1i})}.\nonumber
 \end{eqnarray}}%
By the Lyapunov inequality and the above inequality, we have
{\setlength\abovedisplayskip{1pt}
 \setlength\belowdisplayskip{1pt}
 \begin{eqnarray}\label{vlpn15}
\liminf\limits_{t\rightarrow \infty}\left(\mathbb{E}\left[|y_i(t)-y_0(t)|^2\right]\right)^{\frac{1}{2}}
\geqslant\liminf\limits_{t\rightarrow \infty}\mathbb{E}\left[|y_i(t)-y_0(t)|\right]
\geqslant\frac{|c_i b_i k_{2i}k^2\sigma_{i0}\Upsilon_{i0}c_0|}{(a_0-kc_0)(a_i+b_ik_{1i})},\nonumber
 \end{eqnarray}}%
which implies
$\liminf\limits_{t\rightarrow \infty}\mathbb{E}\left[|y_i(t)-y_0(t)|^2\right]
\geqslant\frac{c_i^2 b_i^2 k_{2i}^2k^4\sigma_{i0}^2\Upsilon^2_{i0}c^2_0}{(a_0-kc_0)^2(a_i+b_ik_{1i})^2}.$


\vspace{0.2cm}
\emph{Proof of Theorem \ref{theorem4}}:
Denote $e_i(t)$ $=x_i(t)-\hat{x}_i(t)$ and
$\delta_i(t)=\hat{x}_{i0}(t)-x_0(t)$, $i=1,\ldots N$.
By $(\ref{b1})$ and $(\ref{b3})$, we  get
{\setlength\abovedisplayskip{4pt}
\setlength\belowdisplayskip{4pt}
  \begin{eqnarray}\label{oplpn1}
e_i(t)=e^{\left(a_i-h_i c_i\right)t}e_i(0),\quad i=1,\ldots N.\nonumber
\end{eqnarray}}%
By $(\ref{b2})$, $(\ref{b4})$ and Assumption $\ref{Ak2}$, we have
{\setlength\abovedisplayskip{4pt}
\setlength\belowdisplayskip{4pt}
  \begin{eqnarray}
\textrm{d}\delta_i(t)
=\left(a_0-G_{2i}c_0\right)\delta_i(t)\textrm{d}t-G_{2i}\sigma_{i0}c_0\delta_i(t)\textrm{d}w_{2i0}(t),\nonumber
\end{eqnarray}}%
which implies
{\setlength\abovedisplayskip{4pt}
\setlength\belowdisplayskip{4pt}
  \begin{eqnarray}
\textrm{d}\delta_i(t)
=\left(a_0-G_{2i}c_0\right)\delta_i(t)\textrm{d}t-G_{2i}\sigma_{i0}c_0\delta_i(t)\textrm{d}w_{2i0}(t).\nonumber
\end{eqnarray}}%
By Assumption $\ref{Ak1}$ and the above equation, we get
{\setlength\abovedisplayskip{4pt}
\setlength\belowdisplayskip{4pt}
  \begin{eqnarray}
\mathbb{E}\left[|\delta_i(t)|^2\right]
=e^{\left (2a_0-2G_{2i}c_0+G_{2i}^2\sigma_{i0}^2c_0^2\right)t}\mathbb{E}\left[|\delta_i(0)|^2\right].\nonumber
\end{eqnarray}}%
By the above equation,  we know that there exist  constants $G_{2i}$, $i=1,\ldots N$ such that
$\lim\limits_{t\rightarrow \infty}\mathbb{E}[|\delta_i(t)|^2]=0$, $i=1,\ldots N$, if and only if there exist  constants $G_{2i}$, $i=1,\ldots N$ such that
{\setlength\abovedisplayskip{4pt}
\setlength\belowdisplayskip{4pt}
  \begin{eqnarray}\label{mlplvlpn20}
2a_0-2G_{2i}c_0+G_{2i}^2\sigma_{i0}^2c_0^2<0,\quad i=1,\ldots N,
\end{eqnarray}}%
which is equivalent to  $\sigma^2a_0<\frac{1}{2}$,  where $\sigma^2= \max\limits_{1\leqslant i\leqslant N} \sigma_{i0}^2$.

Sufficiency: If $\sigma^2a_0<\frac{1}{2}$, by $(\ref{mlplvlpn20})$, we choose $G_{2i}$ such that $2a_0-2G_{2i}c_0+G_{2i}^2\sigma_{i0}^2c_0^2<0$, $i=1,\ldots N$. By $b_i\neq 0$  and $c_i\neq 0,$
we choose $k_{1i}$ and $h_i$ such that $a_i+b_ik_{1i}<0$ and $a_i-h_i c_i<0$, $i=1,\ldots N$ and choose $k_{2i}=\gamma_i-k_{1i}\pi_i$, $i=1,\ldots N$, where  $\pi_i=\frac{c_0}{c_i}$ and $\gamma_i=\frac{a_0c_0-a_ic_0}{b_ic_i}$.
Similar to the proof of  Theorem  $\ref{theorem3}$, we obtain
{\setlength\abovedisplayskip{1pt}
\setlength\belowdisplayskip{1pt}
\begin{eqnarray}
\lim\limits_{t\rightarrow \infty}\mathbb{E}\left[\left| y_i(t)-y_0(t)\right|^2\right]=0, \quad i=1, \ldots, N. \nonumber
\end{eqnarray}}%
Necessity:  We use the reduction to absurdity.
 If $\sigma^2a_0\geqslant\frac{1}{2}$, there don't exist  constants $G_{2i}$, $i=1,\ldots N$  such that $(\ref{mlplvlpn20})$ holds.  Therefore, we get $\liminf\limits_{t\rightarrow \infty}\mathbb{E}[|\delta_i(t)|^2]>0$ for any $G_{2i}\in\mathbb{R}$, $i=1,\ldots N$.

Denote $\vartheta_i(t)=y_i(t)-y_0(t),i=1,\ldots,N$.
Noting that $k_{2i}=\gamma_i-k_{1i}\pi_i,i=1,\ldots,N$, by $(\ref{b2})$, $(\ref{b1})$ and $(\ref{1})$, we have
{\setlength\abovedisplayskip{1pt}
  \setlength\belowdisplayskip{1pt}
   \begin{eqnarray}
\dot{\vartheta}_i(t)
=&&\hspace{-.7cm}c_i\dot{x}_i(t)-c_0\dot{x}_0(t)\nonumber\\
=&&\hspace{-.7cm}c_i\left(a_ix_i(t)+b_i u_i(t)-\pi_i a_0x_0(t)\right)\nonumber\\
=&&\hspace{-.7cm}a_i y_i(t)+c_i b_i\left(k_{1i}\hat{x}_{i}(t)+ k_{2i}\hat{x}_{i0}(t)\right)-c_i\left(a_i\pi_i+b_i\gamma_i \right)x_0(t)\nonumber\\
=&&\hspace{-.7cm} \left({a_i}+b_{i}k_{1i}\right)\vartheta_i(t) +b_i k_{1i}\left(\hat{y}_{i}(t)-y_i(t)\right)
+b_i k_{2i}\left(\hat{y}_{i0}(t)-y_0(t)\right).\nonumber
 \end{eqnarray}}%
Integrating both sides of  the above equation from $0$ to $t$, by the definition of $\hat{y}_{i}(t)-y_i(t)$ and $\hat{y}_{i0}(t)-y_0(t)$, we obtain
 {\setlength\abovedisplayskip{1pt}
 \setlength\belowdisplayskip{1pt}
 \begin{eqnarray}\label{lpvoplpn10}
\vartheta_i(t)
=&&\hspace{-.7cm}e^{(a_i+b_i k_{1i})t}\vartheta_i(0)+\int_0^te^{(a_i+b_ik_{1i})(t-s)}b_i k_{1i}\left(\hat{y}_{i}(s)-y_i(s)\right)\textrm{d}s\nonumber\\
&&\hspace{-.7cm}+\int_0^te^{(a_i+b_ik_{1i})(t-s)}b_i k_{2i}
\left(\hat{y}_{i0}(s)-y_0(s)\right)\textrm{d}s\nonumber\\
=&&\hspace{-.7cm}e^{(a_i+b_i k_{1i})t}\vartheta_i(0)-\int_0^te^{(a_i+b_ik_{1i})(t-s)}b_i k_{1i}c_i
e^{\left(a_i-h_i c_i\right)s}e_i(0)\textrm{d}s\nonumber\\
&&\hspace{-.7cm}+\int_0^te^{(a_i+b_ik_{1i})(t-s)}b_i k_{2i}c_0 e^{\left(a_0-G_{2i}c_0-\frac{1}{2}G_{2i}^2\sigma_{i0}^2c_0^2\right)s-G_{2i}\sigma_{i0}c_0w_{2i0}(s)}\delta_i(0)\textrm{d}s.
 \end{eqnarray}}%
 Case (i): The gain $k_{1i}$  satisfies $a_i+b_ik_{1i}> 0$.

 By the definitions of $\vartheta_i(t)$, $e_i(t)$ and $\delta_i(t)$, we know that  there exist initial values
 $x_0(0)$, $x_i(0)$,  $\hat{x}_i(0)$ and $\hat{x}_{i0}(0)$ such that $x_i(0)\neq \frac{c_0x_0(0)}{c_i}$, $\hat{x}_i(0)=x_i(0)$  and $\hat{x}_{i0}(0)=x_0(0)$, i.e. $\vartheta_i(0)\neq 0$, $e_i(0)=0$ and $\delta_i(0)=0$, $i=1,\ldots,N$.

 By $\vartheta_i(t)\neq 0$, $e_i(t)=0$, $\delta_i(t)=0$ and $(\ref{lpvoplpn10})$, we get
 {\setlength\abovedisplayskip{1pt}
 \setlength\belowdisplayskip{1pt}
 \begin{eqnarray}
\vartheta_i(t)=e^{(a_i+b_i k_{1i})t}\vartheta_i(0),\nonumber
 \end{eqnarray}}%
 which implies
 {\setlength\abovedisplayskip{1pt}
 \setlength\belowdisplayskip{1pt}
 \begin{eqnarray}
\mathbb{E}\left[\left|\vartheta_i(t)\right|^2\right]=e^{2\left(a_i+b_i k_{1i}\right)t}\mathbb{E}\left[\left|\vartheta_i(0)\right|^2\right].\nonumber
 \end{eqnarray}}%
Combing $a_i+b_ik_{1i}> 0$, the above equation and the definition of $\vartheta_i(t)$, we obtain
{\setlength\abovedisplayskip{1pt}
 \setlength\belowdisplayskip{1pt}
 \begin{eqnarray}
\lim\limits_{t\rightarrow \infty}\mathbb{E}\left[\left|y_i(t)-y_0(t)\right|^2\right]=\infty,\quad i=1,\ldots,N.\nonumber
 \end{eqnarray}}%
Case (ii): The gain $k_{1i}$  satisfies $a_i+b_ik_{1i}=0$.

  Similar to the case (i), we obtain
 {\setlength\abovedisplayskip{1pt}
 \setlength\belowdisplayskip{1pt}
 \begin{eqnarray}
\mathbb{E}\left[\left|\vartheta_i(t)\right|^2\right]=\mathbb{E}\left[\left|\vartheta_i(0)\right|^2\right],\nonumber
 \end{eqnarray}}
 which together with  $\vartheta_i(0)\neq 0$ gives
 {\setlength\abovedisplayskip{1pt}
 \setlength\belowdisplayskip{1pt}
 \begin{eqnarray}
\lim\limits_{t\rightarrow \infty}\mathbb{E}\left[\left|\vartheta_i(t)\right|^2\right]=\mathbb{E}\left[\left|\vartheta_i(0)\right|^2\right]>0.\nonumber
 \end{eqnarray}}%
 By the above equation and the definition of $\vartheta_i(t)$, we have
 {\setlength\abovedisplayskip{1pt}
 \setlength\belowdisplayskip{1pt}
 \begin{eqnarray}
\lim\limits_{t\rightarrow \infty}\mathbb{E}\left[\left|y_i(t)-y_0(t)\right|^2\right]=\mathbb{E}\left[\left|\vartheta_i(0)\right|^2\right]>0,\quad i=1,\ldots,N.\nonumber
 \end{eqnarray}}%
Case (iii): The gain $k_{1i}$  satisfies $a_i+b_ik_{1i}<0$.

  By the definitions of $\vartheta_i(t)$, $e_i(t)$ and $\delta_i(t)$, we know that  there exist initial values
 $x_0(0)$, $x_i(0)$,  $\hat{x}_i(0)$ and $\hat{x}_{i0}(0)$ such that $x_i(0)= \frac{c_0x_0(0)}{c_i}$, $\hat{x}_i(0)=x_i(0)$  and $\hat{x}_{i0}(0)\neq x_0(0)$, i.e. $\vartheta_i(0)= 0$, $e_i(0)=0$ and $\delta_i(0)\neq 0$, $i=1,\ldots,N$.

 Noting that  $\vartheta_i(0)= 0$, $e_i(0)=0$ and $\delta_i(0)\neq 0$, $i=1,\ldots,N$, by $(\ref{lpvoplpn10})$, we get
 {\setlength\abovedisplayskip{1pt}
 \setlength\belowdisplayskip{1pt}
 \begin{eqnarray}
\vartheta_i(t)
=\int_0^te^{(a_i+b_ik_{1i})(t-s)}b_i k_{2i}c_0e^{\left(a_0-G_{2i}c_0-\frac{1}{2}G_{2i}^2\sigma_{i0}^2c_0^2\right)s-G_{2i}\sigma_{i0}c_0w_{2i0}(s)}\delta_i(0)\textrm{d}s.\nonumber
 \end{eqnarray}}%
 Taking the absolute value and
mathematical expectation  of the above equation, we have
{\setlength\abovedisplayskip{1pt}
\setlength\belowdisplayskip{1pt}
\begin{eqnarray}\label{plvlpn4}
\mathbb{E}\left[\left| \vartheta_i(t)\right|\right]=&&\hspace{-.7cm}|b_i k_{2i}c_0| \int_0^t
\mathbb{E}\Big[\Big|e^{\left(a_0-G_{2i}c_0-\frac{1}{2}G_{2i}^2\sigma_{i0}^2c_0^2\right)s-G_{2i}\sigma_{i0}c_0w_{2i0}(s)}\delta_i(0)\Big|\Big]
 e^{(a_i+b_ik_{1i})(t-s)}\textrm{d}s\nonumber\\
=&&\hspace{-.7cm}|b_i k_{2i}c_0| \int_0^t
e^{(a_i+b_ik_{1i})(t-s)}\mathbb{E}\left[\left|\delta_i(s)\right|\right] \textrm{d}s.
\end{eqnarray}}
Noting that $\liminf\limits_{t\rightarrow \infty}\mathbb{E}[|\delta_i(t)|]>0$, we know that there exists a constant $c>0$ such that $\liminf\limits_{t\rightarrow \infty}\mathbb{E}[|\delta_i(t)|]$\\$=c$.
Therefore, for any given constant $0<\varepsilon<c$, there exists a positive constant $T>0$ such that for any $t>T$,
$\mathbb{E}[|\delta_i(t)|]>c-\varepsilon$, which gives
{\setlength\abovedisplayskip{1pt}
\setlength\belowdisplayskip{1pt}
\begin{eqnarray}\label{plvlpn8}
\liminf\limits_{t\rightarrow \infty}\int_T^t
e^{(a_i+b_ik_{1i})(t-s)}\mathbb{E}\left[\left|\delta_i(s)\right|\right] \textrm{d}s
\geqslant&&\hspace{-.7cm}\frac{c-\varepsilon}{-(a_i+b_i k_{1i})}\lim\limits_{t\rightarrow \infty}\left[1-e^{(a_i+b_ik_{1i})(t-T)}\right]\nonumber\\
=&&\hspace{-.7cm}\frac{c-\varepsilon}{-(a_i+b_i k_{1i})}>0.
\end{eqnarray}}%
By $a_i+b_ik_{1i}<0$, we get
{\setlength\abovedisplayskip{1pt}
\setlength\belowdisplayskip{1pt}
\begin{eqnarray}\label{plvlpn9}
\lim\limits_{t\rightarrow \infty}\int_0^Te^{(a_i+b_ik_{1i})(t-s)}\mathbb{E}\left[\left|\delta_i(s)\right|\right] \textrm{d}s
=\lim\limits_{t\rightarrow \infty}e^{(a_i+b_ik_{1i})t}\int_0^Te^{-(a_i+b_ik_{1i})s}\mathbb{E}\left[\left|\delta_i(s)\right|\right] \textrm{d}s
=0,\nonumber
\end{eqnarray}}%
which together with  $(\ref{plvlpn8})$ gives
{\setlength\abovedisplayskip{1pt}
\setlength\belowdisplayskip{1pt}
\begin{eqnarray}\label{hgplvlpn10}
\liminf\limits_{t\rightarrow \infty}\int_0^t
e^{(a_i+b_ik_{1i})(t-s)}\mathbb{E}\left[\left|\delta_i(s)\right|\right] \textrm{d}s>0,\quad i=1,\ldots,N.
\end{eqnarray}}%
By $\sigma^2a_0\geqslant\frac{1}{2}$, we have $a_0>0$. By $k_{2i}=\gamma_i-k_{1i}\pi_i$, $\pi_i=\frac{c_0}{c_i}$ and $\gamma_i=\frac{a_0c_0-a_ic_0}{b_ic_i}$, $i=1,\ldots N$, we get
$$|b_i k_{2i}c_0|=|b_ic_0||k_{2i}|=|b_ic_0|\left|\frac{\left(a_0-(a_i+b_i k_{1i})\right)c_0}{b_ic_i}\right|=|b_ic_0|\frac{\left|a_0-(a_i+b_i k_{1i})\right||c_0|}{|b_ic_i|},$$
which together with $a_i+b_i k_{1i}<0$, $b_i\neq0$, $c_i\neq0$, $i=1, \ldots, N$, $a_0> 0$ and $c_0\neq0$ gives
$|b_i k_{2i}c_0|>0$.

Combining $(\ref{plvlpn4})$, $(\ref{hgplvlpn10})$, $|b_i k_{2i}c_0|>0$ and the definition of $\vartheta_i(t)$, we obtain
{\setlength\abovedisplayskip{1pt}
\setlength\belowdisplayskip{1pt}
\begin{eqnarray}
\liminf\limits_{t\rightarrow \infty}\mathbb{E}\left[\left| y_i(t)-y_0(t)\right|\right]>0, \quad i=1, \ldots, N. \nonumber
\end{eqnarray}}%
By the Lyapunov inequality and the above inequality, we have
{\setlength\abovedisplayskip{1pt}
 \setlength\belowdisplayskip{1pt}
 \begin{eqnarray}\label{vlpn15}
\liminf\limits_{t\rightarrow \infty}\left(\mathbb{E}\left[|y_i(t)-y_0(t)|^2\right]\right)^{\frac{1}{2}}
\geqslant&&\hspace{-.7cm}\liminf\limits_{t\rightarrow \infty}\mathbb{E}\left[|y_i(t)-y_0(t)|\right]>0,,\quad i=1,\ldots,N,\nonumber
 \end{eqnarray}}%
 which implies
 $\liminf\limits_{t\rightarrow \infty}\mathbb{E}\left[|y_i(t)-y_0(t)|^2\right]>0$,\quad i=1,\ldots,N.

In summary,  for any $G_{2i}\in\mathbb{R}$, $k_{1i}\in\mathbb{R}$ and $h_i\in\mathbb{R}$, $i=1, \ldots, N$, we know that the HMASs $(\ref{b2})$$-$$(\ref{b1})$ can't achieve mean square  output tracking, which is contrary to  the fact that there exists an admissible observation strategy $J\in\mathcal{J}$ and an admissible cooperative control strategy  $U\in\mathcal{U}$ such that  the HMASs $(\ref{b2})$$-$$(\ref{b1})$ achieve mean square  output tracking.
Therefore,  we have $\sigma^2a_0<\frac{1}{2}$.
 $\hfill\square$


\begin{thebibliography}{99}

\bibitem{Olfati-Saber}
 R. Olfati-Saber and R. M. Murray,  ``Consensus problems in networks of agents with switching topology and time-delays,'' \emph{IEEE Trans. Automat. Control}, vol. 49,  no.   9, pp. 1520$-$1533, Sep. 2013.


\bibitem{Ren213}
 W. Ren and R. W. Beard,  ``Consensus seeking in multiagent systems under
dynamically changing interaction topologies,'' \emph{IEEE Trans. Automat. Control}, vol. 50, no. 5, pp. 655$-$661, May 2005.


\bibitem{Salehi}
 A. Tahbaz-Salehi and A. Jadbabaie,  ``A necessary and sufficient condition for consensus over random networks,''
 \emph{IEEE Trans. Automat. Control}, vol. 53, no. 3, pp. 791$-$795, Apr. 2008.







\bibitem{SuSZ2}  S. Su and  Z. Lin,
``Distributed consensus control of multi-agent systems with higher order agent dynamics and dynamically
changing directed interaction topologies,'' \emph{IEEE Trans. Automat. Control}, vol. 61, no. 2, pp. 515$-$519,
Feb. 2016.



\bibitem{ZhiyongYu}  Z. Yu, S. Yu, H.  Jiang, and C. Hu,
``Observer-based consensus for multi-agent systems with partial adaptive dynamic protocols,''
\emph{Nonlinear Anal. Hybri.}, vol. 34, pp. 58$-$73, Nov. 2019.


%

%
%
%
%
%


\bibitem{Brouwer} D. Brouwer,
``Solution of the problem of artificial satellite theory without drag,'' \emph{Astron. J.},  vol. 64,
 pp. 378$-$396, Nov. 1959.




\bibitem{Murray2}   R. M. Murray, S. S. Sastry, and  L. Zexiang,
 \emph{A mathematical introduction to robotic manipulation},
Florida, USA: CRC Press, 1994.






\bibitem{Bevrani}   H.  Bevrani, \emph{ Real power compensation and frequency control}, New York, USA: Springer, 2009.


\bibitem{Wieland} P. Wieland and F. Allg${\rm\ddot{o}}$wer,
``An internal model principle for consensus in heterogeneous linear multi-agent systems,'' in \emph{ Proc. 1st IFAC Workshop Estim. Control Netw. Syst.}, Venice, Italy, Sep.
2009, pp. 7$-$12.


\bibitem{Francis}  B. A.  Francis and W. M. Wonham,
 ``The internal model principle of control theory,''  \emph{Automatica}, vol. 12, no. 5, pp. 457$-$465, Sep. 1976.

\bibitem{Wieland1}  P. Wieland, R. Sepulchre, and F. Allg${\rm\ddot{o}}$wer,
``An internal model principle is necessary and sufficient for linear output synchronization,'' \emph{Automatica}, vol. 47, no.  5, pp. 1068$-$1074, May 2011.


\bibitem{Lunze} J. Lunze,
``Synchronization of heterogeneous agents,'' \emph{IEEE Trans. Automat. Control}, vol. 57, no.  11, pp. 2885$-$2890,  Nov. 2012.


\bibitem{Grip}  H. F. Grip, T. Yang, A. Saberi, and A. A. Stoorvogel,
``Output synchronization for heterogeneous networks of
non-introspective agents,'' \emph{ Automatica},  vol. 48, no.  10, pp. 2444$-$2453,   Oct. 2012.


\bibitem{LewisF.L.66} F. L. Lewis, B.  Cui, T.  Ma, Y.  Song, and C.  Zhao,
``Heterogeneous multi-agent systems: reduced-order
synchronization and geometry,''   \emph{IEEE Trans. Automat. Control}, vol. 61, no.  5, pp. 1391$-$1396,   May 2016.


\bibitem{Francis1}   B. A. Francis,
``The linear multivariable regulator problem,''  \emph{SIAM J. Control Optim.}, vol. 15, no. 3, pp. 486$-$505, 1977.

\bibitem{Huang}  J. Huang and W. J. Rugh,
``On a nonlinear multivariable servomechanism problem,''  \emph{Automatica}, vol. 26, no.  6, pp. 963$-$972, Nov. 1990.



\bibitem{Su}  Y.  Su and J. Huang,
 ``Cooperative output regulation of linear multi-agent systems,'' \emph{IEEE Trans. Automat. Control},
  vol. 57, no.  4, pp. 1062$-$1066, Apr. 2012.



 \bibitem{Alvergue}   L. D. Alvergue, A. Pandey, G.  Gu, and X. Chen,
 ``Consensus control for heterogeneous multiagent systems,''  \emph{SIAM J. Control Optim.},
  vol. 54, no.  3, pp. 1719$-$1738, Jun. 2016.



 \bibitem{Yan43} F. Yan, G.  Gu,  and X. Chen,
``A new approach to cooperative output regulation for heterogeneous multi-agent systems,'' \emph{SIAM J. Control Optim.},
  vol. 56, no.  3, pp. 2074$-$2094,  Mar.  2018.



 \bibitem{Huang74} C.  Huang and X. Ye, ``Cooperative output regulation of heterogeneous
multi-agent systems: an $H_{\infty}$ criterion,''  \emph{IEEE Trans. Automat. Control},
  vol. 59, no.  1, pp. 267$-$273,   Jan. 2014.

\bibitem{Yaghmaie74}  F. A. Yaghmaie, F. L. Lewis, and R. Su, ``Output regulation of linear heterogeneous multi-agent systems via output and state feedback,'' \emph{Automatica},  vol. 67,  pp. 157$-$164, May 2016



 \bibitem{Meng74}  Z.  Meng, T. Yang, D. V. Dimarogonas, and K. H. Johansson,
``Coordinated output regulation of heterogeneous linear systems under
switching topologies,'' \emph{Automatica},  vol. 53,  pp. 362$-$368, Mar. 2015.
\bibitem{Kim}  H. Kim, H. Shim, and J. H. Seo,
``Output consensus of heterogeneous uncertain linear multi-agent systems,''
 \emph{IEEE Trans. Automat. Control}, vol. 56, no. 1, pp. 200$-$206,  Jan. 2011.






\bibitem{Su1}  Y.  Su, Y.  Hong, and J. Huang,
``A general result on the robust cooperative output
regulation for linear uncertain multi-agent systems,'' \emph{IEEE Trans. Automat. Control}, vol. 58, no. 5, pp. 1275$-$1279,  May 2013.




%
%
%
%


\bibitem{DingZT42}  Z. Ding,
 ``Consensus output regulation of a class of heterogeneous nonlinear systems,''
  \emph{IEEE Trans. Automat. Control}, vol.  58, no.  10,  pp. 2684$-$2653,  Oct. 2013.







 \bibitem{WangXH420}
 X.  Wang, Y.  Su, and D.  Xu, ``A nonlinear internal model design for heterogeneous second-order
multi-agent systems with unknown leader,''  \emph{Automatica}, vol.  91,  pp. 27$-$35, May 2018.






%

\bibitem{Huang88}
  M.  Huang and J. H. Manton, ``Coordination and consensus of networked agents with noisy measurement: stochastic algorithms and asymptotic behavior,''  \emph{SIAM J. Control Optim.}, vol. 48, no. 1, pp. 134$-$161, Feb. 2009.



\bibitem{Aysal10}   T. C.  Aysal and K. E. Barner,  ``Convergence of consensus models with stochastic disturbances,''
\emph{IEEE Trans. Automat. Control}, vol. 56, no. 8, pp. 4101$-$4113, Aug. 2010.



\bibitem{Kar0}  S. Kar  and J. M.  Moura,  ``Distributed consensus algorithms in sensor networks with imperfect communication: link failures and channel noise,'' \emph{IEEE T. Signal Proces.}, vol. 57, no. 1, pp. 355$-$369, Jan. 2009.


\bibitem{Li11} T. Li  and J. F. Zhang,
 ``Consensus conditions of multi-agent systems with time-varying topologies and stochastic communication noises,''
\emph{IEEE Trans. Automat. Control}, vol. 55, no. 9, pp. 2043$-$2057, Sep. 2010.



\bibitem{Huang11}    M. Huang, T. Li, and J. F. Zhang,
``Stochastic approximation based consensus dynamics over markovian networks,''
\emph{SIAM J. Control Optim.},
vol. 53, no. 6, pp. 3339$-$3363, Nov. 2015.


\bibitem{Huang55}  M.   Huang and  J. H. Manton,
 ``Stochastic consensus seeking with noisy and directed
inter-agent communication: fixed and randomly varying topologies,'' \emph{IEEE Trans. Automat. Control},
vol. 55, no. 1, pp. 235$-$241, Jan. 2010.



\bibitem{Mac12}    C.  Ma, T. Li, and J.  Zhang,
 ``Consensus control for leader-following multi-agent systems with measurement noises,''
 \emph{J. Syst. Sci. Complex.}, vol. 23, no. 1, pp. 35$-$49, Feb. 2010.


 \bibitem{Hu59}  J.  Hu and G. Feng,
``Distributed tracking control of leader-follower multi-agent systems under
noisy measurement,''  \emph{Automatica}, vol. 46, no.  8, pp. 1382$-$1387, Aug. 2010.




\bibitem{ZhengYS23}    Y.  Zheng, W. S Chen, and L. Wang,
``Finite-time consensus for stochastic multi-agent systems,''
\emph{ Int. J. Control},  vol. 84, no.  10, pp. 1644$-$1652, Sep. 2011.


\bibitem{LiW.Q.22} W.  Li, L.  Xie, and J. F. Zhang,
``Containment control of leader-following multi-agent systems with
Markovian switching network topologies and measurement noises,'' \emph{Automatica},
vol. 51,  pp. 263$-$267, Jan. 2015.





\bibitem{Liu22}   J. Liu, X.  Liu, W. C. Xie, and H.  Zhang,
``Stochastic consensus seeking with communication delays,''
\emph{Automatica}, vol. 47, no. 12, pp. 2689$-$2696, Dec. 2011.

\bibitem{Cheng22} L. Cheng, Z. G. Hou, M. Tan, and X. Wang,
``Necessary and sufficient conditions for consensus
of double-integrator multi-agent systems with measurement noises,''
\emph{IEEE Trans. Automat. Control}, vol. 56, no. 8, pp. 1958$-$1963, Aug. 2011.

\bibitem{LiuX.1}  X.  Liu, B.  Xu, and L.  Xie, ``Distributed tracking control of second-order multi-agent systems under
measu-rement noises,'' \emph{J. Syst. Sci. Complex}, vol. 27, no. 5, pp. 853$-$865, Apr. 2014.


\bibitem{LiW.Q.}   W.  Li, T. Li, L. Xie, and J. F. Zhang,
 ``Necessary and sufficient conditions for bounded distributed mean
square tracking of multi-agent systems with noises,''  \emph{Int. J. Robust Nonlin.},
vol. 26, no. 4, pp. 631$-$645,  Mar. 2016.



\bibitem{WuZ.H.}  Z.  Wu, L. Peng,  L.  Xie, and J.  Wen,
 ``Stochastic bounded consensus tracking of second-order multi-agent systems with measurement
noises based on sampled-data with general sampling delay,''   \emph{Int. J. Syst. Sci.},
vol. 46, no.  3, pp. 546$-$561,  Feb. 2015.


\bibitem{Cheng32}  L. Cheng, Z. G. Hou, and M. Tan,
``A mean square consensus protocol for linear multi-agent
systems with communication noises and fixed topologies,'' \emph{IEEE Trans. Automat. Control},
vol. 59, no.     1, pp. 261$-$267,  Jan. 2014.
\bibitem{Huang213}  L.  Huang, H. Hjalmarsson, and H. Koeppl,
``Almost sure stability and stabilization of discrete-time stochastic systems,''
 \emph{Syst. Control Lett.},  vol. 82, pp. 26$-$32,  Aug. 2015.

%
%
%
%
%


 \bibitem{Ni22}    Y. H. Ni and X. Li,
   ``Consensus seeking in multi-agent systems with multiplicative measurement
noises,''  \emph{Syst. Control Lett.}, vol. 62, no. 5, pp. 430$-$437,
May 2013.




\bibitem{Djaidja22}  S. Djaidja and Q.   Wu,
  ``Leader-following consensus for single-integrator multi-agent systems with multiplicative
noises in directed topologies,''  \emph{Int. J. Syst. Sci.}, vol. 46, no. 15, pp. 2788$-$2798,
Nov. 2015.





\bibitem{Li1}   T. Li, F. Wu, and J. F. Zhang,
 ``Multi-agent consensus with relative state-dependent measurement noises,''
    \emph{IEEE Trans. Automat. Control}, vol. 59, no. 9, pp. 2463$-$2468,
Sep. 2014.

\bibitem{Zong2}  X.  Zong, T. Li, and J. F. Zhang,
 ``Consensus conditions of continuous-time multi-agent systems with additive and multiplicative measurement noises,''
   \emph{SIAM J. Control Optim.}, vol. 56, no. 1, pp. 19$-$52, Jan. 2018.


\bibitem{Zong33} X.  Zong, T. Li, and J. F. Zhang,
``Consensus conditions of continuous-time multi-agent systems with
time-delays and measurement noises,'' \emph{Automatica}, vol. 99, pp. 412$-$419, Jan. 2019.




\bibitem{Zong586}
X. F. Zong, T. Li, and J. F. Zhang, ``Stochastic consensus of linear multi-agent systems with multiplicative
measurement noises,''
in  \emph{ Proc. 12th IEEE Intl. Conf.  Control  Automat.},  Kathmandu, Nepal, Jun. 2016, pp. 7-12.


\bibitem{Zong5869}   X. Zong, T. Li, G. Yin, and J. F. Zhang,
``Stochastic consentability of linear systems with time delays and multiplicative noises,''
\emph{IEEE Trans. Automat. Control}, vol. 63, no.  4, pp. 1059$-$1074, Apr. 2018.




  \bibitem{Wangt22}   J. Wang, and N. Elia, ``Distributed averaging under constraints on information exchange: emergence of L${\rm \acute{e}}$vy flights,''  \emph{ IEEE Trans. Automat. Control}, vol. 57, no. 10,  pp. 2435$-$2449, Oct.  2012.

 \bibitem{Wangt23}   J. Wang, and N. Elia, `` Mitigation of complex behavior over networked systems: analysis of spatially invariant structures,''
  \emph{ Automatica}, vol. 49, no. 6,  pp. 1626$-$1638, Jun.  2012.

\bibitem{Fourati12}   H. Fourati,
 \emph{Multi-sensor Data Fusion: From Algorithms and Architectural Design to Applications},
  Boca Raton, USA: CRC Press, 2015.
\bibitem{KSchertzer}
D. Schertzer, M. Larcheveque, J. Duan, V. V. Yanovsky, and S. Lovejoy,
``Fractional Fokker-Planck equation for nonlinear stochastic differential equations driven by non-Gaussian L\'{e}vy stable noises,'' \emph{ J. Math. Phys.}, vol. 42, no. 1,  pp. 200$-$212, Jan. 2001.

\bibitem{Bco2} L. J. Bo, K. H. Shi, and Y. J. Wang,
`` On a stochastic wave equation driven by a non-gaussian L\'{e}vy process,'' \emph{ J. Theor. Probab.}, vol. 23,  pp. 328$-$343, Mar. 2010.


\bibitem{KGaoi42} T. Gao and J. Duan,
``Quantifying model uncertainty in dynamical systems driven by non-Gaussian L\'{e}vy stable noise with observations on mean exit time or escape probability,'' \emph{ Commun. Nonlinear Sci. Numer. Simulat.}, vol. 39,  pp. 1$-$6, Oct. 2016.



\bibitem{Karimi2} D. Zhang, Z. H. Xu, H. R. Karimi, Q. G. Wang, and L. Yu, ``Distributed $ H_{\infty} $ output-feedback control for consensus of heterogeneous linear multiagent systems with aperiodic sampled-data communications,''    \emph{IEEE Trans. Ind. Electron.}, vol. 65, no. 5,  pp. 4145$-$4155, May  2018.



\bibitem{Huang12}   J. Huang,
 \emph{Nonlinear Output Regulation: Theory and Applications},
 Philadelphia, USA: SIAM, 2004.









\bibitem{Tomashevich}
S. Khodaverdian,  M. Schneider, and J. Adamy,  ``Synchronizing networks of heterogeneous linear systems via input-output decoupling,''
in \emph{ Proc.  IEEE Conf.  Control Appl.}, Antibes, France, Oct. 2014, pp. 681$-$686.


\bibitem{Mao33}   X. Mao, \emph{Stochastic Differential Equations and Applications},
 Chichester, UK: Horwood, 2006.




%



%
%
%

\end{thebibliography}
\end{document}